\documentclass[aps,pra,twocolumn,notitlepage,amsfonts,citeautoscript,superscriptaddress,showpacs]{revtex4-2}

\usepackage{xfrac}
\usepackage{graphicx,epstopdf,color}
\usepackage{amsmath,amssymb,mathrsfs}
\usepackage{hyperref}
\usepackage{cleveref}
\usepackage{tikz}
\usepackage{float}
\usepackage{placeins}

\crefname{equation}{Eq.}{Eqs.}
\Crefname{equation}{Equation}{Equations}
\crefname{figure}{Fig.}{Figs.}
\Crefname{figure}{Figure}{Figures}
\crefname{section}{Sec.}{Secs.}
\Crefname{section}{Section}{Sections}
\crefname{appendix}{Appendix}{Apps.}
\Crefname{appendix}{Appendix}{Apps.}
\crefname{paragraph}{Sec.}{Secs.}
\crefname{table}{Table}{Tables}

\newcommand{\ket}[1]{\left|#1\right\rangle}
\newcommand{\bra}[1]{\left\langle#1\right|}
\newcommand{\braket}[2]{\bigl\langle#1\bigl|\bigr.#2\bigr\rangle}

\def\ie{i.e.\ }

\graphicspath{{nb_fig/}}

\newcommand{\hn}{\hat{n}}
\newcommand{\hphi}{\hat{\varphi}}
\newcommand{\hH}{\hat{H}}
\newcommand{\ha}{\hat{a}}
\newcommand{\hb}{\hat{b}}
\newcommand{\hc}{\hat{c}}
\DeclareMathOperator{\Tr}{Tr}
\renewcommand{\Im}{\text{Im}}

\makeatletter
\def\@fnsymbol#1{\ensuremath{\ifcase#1\or * \or \mathsection\or \mathparagraph\or \|\or **\or \else\@ctrerr\fi}}

\makeatother

\begin{document}
\title{Flux pump induced degradation of \texorpdfstring{$T_1$}{T1} for dissipative cat qubits}
\author{L{\ifmmode\acute{e}\else\'{e}\fi}on Carde}
\thanks{leon.carde@alice-bob.com}
\affiliation{Laboratoire de Physique de l’École Normale Supérieure, Inria, ENS,
Mines ParisTech, Université PSL, Sorbonne Université, Paris, France}
\affiliation{Alice and Bob, 53 Bd du Général Martial Valin, 75015, Paris, France}
\author{Pierre Rouchon}
\affiliation{Laboratoire de Physique de l’École Normale Supérieure, Inria, ENS,
Mines ParisTech, Université PSL, Sorbonne Université, Paris, France}
\author{Joachim Cohen}
\affiliation{Alice and Bob, 53 Bd du Général Martial Valin, 75015, Paris, France}
\author{Alexandru Petrescu}
\affiliation{Laboratoire de Physique de l’École Normale Supérieure, Inria, ENS,
Mines ParisTech, Université PSL, Sorbonne Université, Paris, France}
\date{\today}
\begin{abstract} 
    Dissipative stabilization of cat qubits autonomously corrects for bit-flip errors by ensuring that reservoir-engineered two-photon losses dominate over other mechanisms inducing phase-flip errors. To describe the latter, we derive an effective master equation for an asymmetrically threaded SQUID based superconducting circuit used to stabilize a dissipative cat qubit. We analyze the dressing of relaxation processes under drives in time-dependent Schrieffer-Wolff perturbation theory for weakly anharmonic bosonic degrees of freedom, and in numerically exact Floquet theory. We find that spurious single-photon decay rates can increase under the action of the parametric pump that generates the required interactions for cat-qubit stabilization. Our analysis feeds into mitigation strategies that can inform current experiments, and the methods presented here can be extended to other circuit implementations. 
\end{abstract}
\maketitle

\section{Introduction}
\label{sec:intro}
Superconducting circuit quantum electrodynamics (cQED) has emerged as one of the leading platforms for quantum information processing due to progress in control, readout, and state preparation~\cite{blais_circuit_2021,kjaergaard_et_al_2019}. However, like all physical systems, superconducting circuits are prone to decoherence~\cite{ithier_et_al_2005}. Bosonic quantum error correction is one way of countering decoherence in cQED, by encoding information redundantly in the infinite Hilbert space of a harmonic oscillator. In particular, the two-legged cat code~\cite{leghtas_hardware-efficient_2013, leghtas_confining_2015} encodes a qubit in the manifold of two coherent states well separated in phase space, thereby offering protection against bit flips caused by local noise~\cite{guillaud_quantum_2023, lescanne_exponential_2020}. Error processes changing the photon number parity, like single photon loss, result in phase-flip errors, and are corrected through classical codes such as the repetition~\cite{guillaud_quantum_2023,putterman_hardware-efficient_2024} or low-density parity check (LDPC) code~\cite{ruiz_ldpc_2025}. For these codes to be operated below threshold, the phase-flip error rate should remain low while increasing the speed at which error correction is performed.

\begin{figure}[t!]
     \centering
    \includegraphics[width=0.95\linewidth]{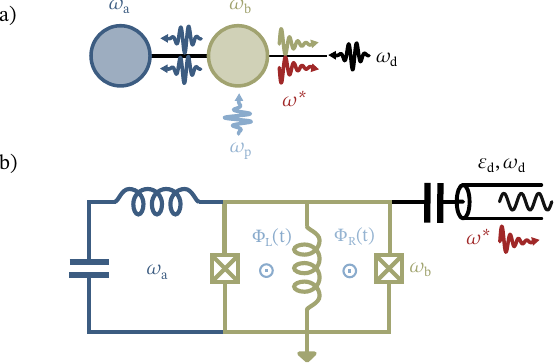}
    \caption{a) Abstract cat-qubit system: a high-Q mode (dark blue) and a non-linear low-Q mode (green) driven to implement a 4-wave coupling (light blue arrow). Wavy lines illustrate parametrically activated 2-1 photon exchange interaction driven at $\omega_p$. Additionally, the `buffer' $b$ mode (green) might be driven through a weak resonant drive (black arrow). Decaying wavy lines on the right represent the strong dissipation of the buffer mode, along with spurious decays in red at frequency $\omega^*$. b) Galvanically coupled circuit for cat-qubit implementation: 
    green branches form the two flux-driven loops of the ATS (two identical Josephson junctions shunted by a superinductance~\cite{lescanne_exponential_2020}), with colors corresponding to a). The dominant dissipation channel of the buffer is represented by a capacitive coupling to a transmission line.
    }
     \label{fig:GalvaCat}
\end{figure}

Dissipative cat qubits autonomously suppress the leakage outside the code manifold preventing bit flips from occurring. In circuit QED, this is realized through a parametric Hamiltonian~\cite{puri_engineering_2017,grimm_stabilization_2020} or by parametrically coupling the cat-qubit resonator (the storage cavity) to a lossy mode (the buffer) to engineer a specific two-photon dissipation~\cite{leghtas_confining_2015,lescanne_exponential_2020}. Our work builds upon the asymmetrically threaded SQUID~\cite{lescanne_escape_2019} (ATS, see \cref{fig:GalvaCat}), which enables a 2-1 photon exchange interaction, where two photons of the cat-qubit mode are swapped with one photon of the buffer. This exchange interaction mediated by the ATS conserves the photon-number parity of the cat-qubit storage cavity, as defined below, and does not affect the phase-flip rate.  However, the constituent Josephson elements of the ATS, when driven, give rise to spurious off-resonant processes which do not conserve the photon-number parity and can result in phase flips~\cite{leghtas_confining_2015,lescanne_exponential_2020}. 

In this work, we investigate the origin of these processes and their impact on cat-qubit devices. Spurious coherent and dissipative mechanisms in the ATS are induced by the interplay of nonlinearity and drives. We provide a quantitative understanding of the parametrically activated processes by deriving an effective master equation in time-dependent Schrieffer-Wolff perturbation theory (SWPT)~\cite{theis_nonadiabatic_2017, petrescu_lifetime_2020,malekakhlagh_first_2020,venkatraman_static_2022} adapted for weakly anharmonic bosonic systems under large-amplitude drives~\cite{petrescu_accurate_2023}, which we then validate with Floquet numerical simulations~\cite{grifoni_driven_1998}. While our approach can apply to multiple circuits with strong parametrically activated interactions facilitating cat-qubit preparation (\cref{fig:GalvaCat}a), we exemplify it on a specific circuit used in recent experiments for dissipative bosonic cat-qubit stabilization (\cref{fig:GalvaCat}b). 

The circuit in \cref{fig:GalvaCat}b)  
consists of two cavities realized as superconducting lumped-element LC circuits coupled galvanically to an ATS. This implements the mode scheme of \cref{fig:GalvaCat}a) in which two harmonic modes, $\ha$ and $\hb$, a memory of high quality factor (dark blue) and a buffer of low quality factor (green), are nonlinearly coupled via a parametrically activated interaction (light blue). With a nonlinear photon exchange interaction between the two modes, the rotating-wave approximation Hamiltonian is~\cite{leghtas_confining_2015}
\begin{align}
    \hH &= g_2\left(\ha^2- \alpha^2 \right)\hb^\dagger + \text{h.c.}, \label{eq:H2pho}
\end{align}
where $g_2$ is the coupling strength increasing with flux pump amplitude, and $g_2\alpha^2$ is the amplitude of the resonant drive on the charge quadrature of the mode $b$ that sets the size of the cat state to $|\alpha|$~\cite{lescanne_exponential_2020}. 

Taking into account the coupling of the modes to external baths, we obtain the Lindblad master equation for dissipative cat state stabilization
\begin{align}
    \mathcal{L}(\hat{\rho})&= -\frac{i}{\hbar} \left[\hH, \hat{\rho} \right] + \kappa_b \mathcal{D}_{\hb} (\hat{\rho})+\kappa_1 \mathcal{D}_{\ha}(\hat{\rho}), \label{eq:calL}
\end{align}
with the dissipator superoperator $\mathcal{D}_{\hat{L}}(\hat{\rho}) = \hat{L}\hat{\rho} \hat{L}^\dagger - \{
\hat{L}^\dagger \hat{L},\hat{\rho} \}/2$ and $\kappa_b,\kappa_1$ the relaxation rates corresponding to single-photon relaxation for the modes $\hb$ and $\ha$, respectively. In the limit where
\begin{align}\label{eq:adiabatic_condition}
    8 g_2 \vert \alpha \vert < \kappa_b,    
\end{align}
the dissipative mode $b$ can be adiabatically eliminated~\cite{azouit_towards_2017, reglade_quantum_2024}, resulting in an effective two-photon driven dissipator on the mode $a$, $\kappa_2 \mathcal{D}_{\hat{a}^2-\alpha^2}(.)$, where $\kappa_2 = 4 g_2^2/\kappa_b$~\cite{mirrahimi_dynamically_2014}. Moreover, combining the adiabatic condition \cref{eq:adiabatic_condition} with this expression for $\kappa_2$ leads to $\kappa_2<\kappa_b/|4\alpha|^2$.

Under the two-photon dissipator, the system is confined to a code space spanned by the cat states, $\ket{\mathcal{C}_{\alpha}^{\pm}} = \mathcal{N}_\pm (\ket{\alpha}\pm\ket{-\alpha})$ with a confinement rate determined by $\kappa_2$. As this rate also sets an upper bound on the speed of the gates required to detect and correct phase-flip errors when concatenating the cat code with a classical code~\cite{guillaud_repet_2019,ruiz_ldpc_2025}, the physical error rate per error correction cycle scales as $\kappa_1/\kappa_2$. For the repetition code to be operated below threshold, one typically requires $\kappa_1/\kappa_2 \lesssim 0.005 $~\cite{guillaud_error_2021, chamberland_building_2022} for a cat size $\alpha = \sqrt{8}$.  Therefore, for such cat sizes, in the ideal case where the adiabaticity condition is respected,  the condition for being under threshold requires the timescale separation 
\begin{align}
\kappa_1 \lesssim 5 \cdot 10^{-3}\kappa_2 \lesssim 4 \cdot 10^{-5} \kappa_b. \label{eq:tsep}
\end{align}

Thus, for the approach above to work, the phase-flip error rate $\kappa_1$ should remain small compared to $\kappa_2$~\cite{leghtas_confining_2015}, and only weakly change under the drives that control the rate $\kappa_2$, despite the large coupling of the system to the environment necessary to induce the buffer loss rate $\kappa_b$. This situation is akin to the degradation of $T_1$ in the dispersive readout of the transmon~\cite{boissonneault_dispersive_2009,petrescu_lifetime_2020, thorbeck_read-out_2024}, where the lossy readout resonator is coupled to a high-Q transmon.  

Here, we address such deviations given by parametrically-activated off-resonant terms, that are not present in the first-order rotating-wave approximation Hamiltonian \cref{eq:H2pho}. 
We classify their contributions in SWPT based on their change under memory-mode parity, $\hat{P}_a \ha \hat{P}_a = -\ha$ where $\hat{P}_a =e^{i\pi \ha^\dagger \ha}$ is the photon-number parity operator. The Hamiltonian $\hat{H}$ in \cref{eq:calL,eq:H2pho} conserves parity, but the Liouvillian $\mathcal{L}$ in \cref{eq:calL} does not (single-photon decay of rate $\kappa_1$). For a cat-qubit-based repetition code, for instance, this induces phase-flips on the logical qubit~\cite{guillaud_repet_2019}. Below, we show that the rate $\kappa_1$ is drive-dependent and that other parity-breaking dissipators, including correlated decay processes between memory and buffer, are generated by off-resonant processes at higher orders of the rotating-wave approximation. All of these changes can become significant for experimentally reasonable parametric pump powers.

The remainder of this paper is organized as follows. In \cref{sec:eff-mod}, we present our drive-amplitude-dependent effective master equation and classify the possible loss channels. We validate SWPT in the strong-drive regime against numerical Floquet simulations in \cref{sec:strong-drive}. Finally, in \cref{sec:mitigation}, we develop a mitigation scheme for spurious decay processes. We conclude in \cref{sec:conclusion}. Technical details are relegated to several appendices as referenced in the text. 

\section{Effective master equation}\label{sec:eff-mod}
In this section, we derive an effective master equation~\cite{petrescu_lifetime_2020,venkatraman_static_2022} for the decay channels of the circuit in  \cref{fig:GalvaCat}. By applying SWPT to both system and system-bath Hamiltonian, we generate and classify all decay processes into the transmission line coupled to the buffer, mediated by the pump on the ATS.  

\subsection{Model Hamiltonian}
The Hamiltonian for the circuit in \cref{fig:GalvaCat}b) is composed of the effective circuit Hamiltonian $\hH_{s}$, and the Hamiltonian $\hH_\textit{sB}$ describing the capacitive coupling of the buffer mode $b$ to the transmission line, whose Hamiltonian is $H_{B}$ (see \cref{app:Norm}), 

\begin{align}\label{eq:H-circuit_1}
\begin{split}
    \frac{\hH_{s}}{\hbar} =& \omega_a \ha^\dagger \ha + \omega_b \hb^\dagger \hb + 2\epsilon_d \cos(\omega_d t )  \left[\hat{y}_b + u \hat{y}_a \right] \\
    &- 2\frac{E_J}{\hbar} \sin\left[\epsilon(t)\right] \sin\left[\varphi_a \hat{x}_a + \varphi_b \hat{x}_b\right],\\
    \frac{\hH_{\textit{sB}}}{\hbar} &=  \left[\hat{y}_b + u\hat{y}_a \right]\otimes \hat{B},\;
    \frac{\hH_B}{\hbar} = \sum_i  \Omega_i \hb_i^\dagger \hb_i.
\end{split}
\end{align}
Here, $\omega_{a/b}$ are the frequencies of the normal modes $a$ and $b$, $E_J$ is the Josephson energy that is common to the two Josephson junctions in \cref{fig:GalvaCat}b), $u$ is the ratio of the hybridization coefficients of the charge operator of the island connected to the transmission line in \cref{fig:GalvaCat}b), $\varphi_a$ and $\varphi_b$ are coefficients encapsulating the impedance of each normal mode, as seen by the junction, defined in \cref{app:Norm}, $\epsilon(t) = \epsilon_p \sin(\omega_pt)$ the flux-pump signal and $\epsilon_d$ the charge-drive amplitude. Numerical values for the coupling constants in \cref{eq:H-circuit_1}, to be used throughout this paper, are given in the caption of \cref{fig:Collapse_monom}. We further introduced $\hb_i$ the bath boson annihilation operators at frequency $\Omega_i$ with the canonical commutation relations $[\hb_i,\hb_j^\dagger]=\delta_{ij}$, and the transmission-line charge operator coupling to the system $\hat{B} = -i \sum_i  g_i (\hb_i - \hb_i^\dagger)$, where $g_i$ are coupling constants with units of energy. Further, we set $\hat{x}_{\eta} = (\hat{\eta} + \hat{\eta}^\dagger )$ and $\hat{y}_{\eta} = (-i)(\hat{\eta} - \hat{\eta}^\dagger )$ for $\eta = a,b$,
the rescaled superconducting phase and Cooper pair number operators corresponding to memory and buffer normal modes.

Turning to the couplings to environmental degrees of freedom, note that, in the typical experimental situation, the largest decay rates are set by the charge coupling to the transmission line in \cref{fig:GalvaCat}b). The direct capacitive coupling of the left node to an external bath is neglected. Moreover, the coupling of the system to the flux lines can result in additional decoherence channels. Nonetheless, we show in \cref{app:bathcoupl} that these contributions are significantly suppressed. 

We then go to the interaction picture with respect to $\hH_0/\hbar =  \frac{\omega_d +\omega_p }{2} \ha^\dagger \ha + \omega_d \hb^\dagger \hb $, where $\omega_p$ stands for the flux-pump frequency while $\omega_d$ is the charge-drive frequency on the buffer mode $b$. This accommodates small detunings of the drives with respect to the resonant conditions required for the cat-state stabilization protocol~\cite{lescanne_exponential_2020}
$\delta = \omega_a - (\omega_p+\omega_d)/2$ and  $\Delta = \omega_b - \omega_d$, defined
with respect to the frequencies of the normal modes  $\omega_{a,b}$ accessible in experiment. 
This allows us to set the frequency of the buffer drive (pump) to the buffer frequency dressed by the drive, leading to a dressed resonance condition necessary to realize the 2-1 photon exchange interaction 
\begin{align}\label{eq:res-cond}
    \begin{split}
        \omega_p &= \vert 2 \tilde{\omega}_a - \tilde{\omega}_b \vert\\
        \omega_d &= \tilde{\omega}_b,
    \end{split}
\end{align}
where $\tilde{\omega}_a, \tilde{\omega}_b$ are the mode frequencies dressed by the pump and buffer drive, as can be obtained to some desired order in perturbation theory, to be defined below. This is analogous to the situation of microwave-activated parametric gates, where the control tone frequency has to be self-consistently matched to the ac Stark shifting frequencies of the targeted states~\cite{malekakhlagh_lifetime_2020,krinner_demonstration_2020,petrescu_accurate_2023}.

After going into the interaction picture in \cref{eq:H-circuit_1}, we make the assumption of small phase oscillations across the ATS due the quadratures of the two bosonic modes. That is, restricting the analysis to states satisfying $ \langle\varphi_a\hat{x}_a+\varphi_b\hat{x}_b\rangle\lesssim1 $, we expand the nonlinear term in the system Hamiltonian $\hH_s$ using both a Taylor expansion and the Jacobi-Anger expansion~\cite{NIST:DLMF} over the harmonics of the flux pump drive $\epsilon(t)$, yielding
\begin{align}
\begin{split} \label{eq:Hs_g_ology}
    \frac{\hH_s(t)}{\hbar} = \delta \ha^\dagger \ha + \Delta \hb^\dagger \hb + 2\epsilon_d \cos(\omega_d t )  [\hat{y}_b(t) + u\hat{y}_a(t) ] \\
    +\sum_{n,k\, \text{odd}} \mathrm{g}_{n,k}\ e^{ik\omega_p t} :\mathrel{\left[\hat{x}_a(t) + r\hat{x}_b(t) \right]^{n}}: +\text{h.c.},
\end{split}
\end{align}
and hereafter use $\hat{H}_s$ to stand for this newly introduced interaction-picture operator, and not  the Schr\"odinger-picture Hamiltonian of \cref{eq:H-circuit_1}.  Here, $:\mathrel{\hat{O}}:$ is the normal-ordered operator $\hat{O}$, $\hat{x}_\eta(t)$ and $\hat{y}_\eta(t)$ the quadratures of the system in the interaction picture for $\eta = a,b$ and we introduced the coupling constants
\begin{align}\label{eq:gs}
    \mathrm{g}_{n,k} = -\frac{2i (-1)^{ \frac{n-1}{2} }}{(n)!} \frac{E_J e^{-\varphi_a^2/2-\varphi_b^2/2}}{\hbar} J_{k}(\epsilon_p) \varphi_a^{n}
\end{align} 
where $J_k$ is the $k$-th Bessel function of the first kind~\cite{NIST:DLMF}, and $r = \frac{\varphi_b }{\varphi_a}$.
Contrary to other versions of SWPT~\cite{petrescu_lifetime_2020,petrescu_accurate_2023,venkatraman_static_2022}, we do not displace~\cite{malekakhlagh_first_2020} the starting Hamiltonian by the classical solutions of the fields corresponding to the charge and phase of the $a$ and $b$ normal modes, since the latter cannot be found analytically. 
We compensate for this drawback by iterating the SWPT to higher orders.

We now introduce notation to track the order of the mixing processes allowed by \cref{eq:Hs_g_ology}. While the angular pump amplitude and the two zero-point fluctuation parameters remain small, \ie $\epsilon_p, \varphi_a, \varphi_b  \ll \pi $, we can use these three parameters to truncate the series in \cref{eq:Hs_g_ology}.
Consistent with experimental values~\cite{reglade_quantum_2024,marquet_autoparametric_2024}, we consider that these three parameters are of the same order of magnitude so that the expansion can be made with respect to a unique small parameter, $\lambda \in \{ \epsilon_p,\varphi_{a,b} \}$ (see \cref{app:trunc}). We further assume $\langle\hat{x}_a \rangle = O(1)$ such that we satisfy the condition $ \langle\varphi_a \hat{x}_a + \varphi_b \hat{x}_b\rangle \lesssim 1 $. Using this notation, the coupling constants introduced above obey $\mathrm{g}_{n,k} = O(\lambda^{n+k})$. Note that a term of \cref{eq:Hs_g_ology} with a prefactor $\lambda^n$ corresponds to an $n$-wave mixing term.

\subsection{Effective master equation}\label{sec:eff-mod:eff-ME}
With these notations, we are ready to proceed to the SWPT. That consists of finding a unitary change of frame in which the system is described by a time-independent effective Hamiltonian
\begin{align}
    \begin{split}\label{eq:Keff}
    \hat{K} - i\hbar \partial_t 
    &\equiv  e^{\hat{S}(t)/i\hbar} [ \hH_s(t) - i\hbar \partial_t ] e^{-\hat{S}(t)/i\hbar} .
    \end{split}
\end{align}
In the above, one expands the effective Hamiltonian $\hat{K}$ and the generator of the transformation $\hat{S}$ with respect to a small parameter of order $O(\lambda^2)$ (for a more detailed discussion see \cref{app:trunc}). The condition that at every order in the small parameter $\lambda$ the effective Hamiltonian remains time-independent results in an equation that is finally solved order by order using the Baker-Campbell-Hausdorff formula~\cite{venkatraman_static_2022, malekakhlagh_first_2020}. Moreover, we truncate the expansion of the Josephson potential in \cref{eq:Hs_g_ology}, in orders of $\lambda$. Applying the SWPT to an already truncated starting point Hamiltonian sets an upper limit to the order of the SWPT that will give contributions consistent with the truncation (\cref{app:trunc}). We therefore refer to the whole procedure solely with the truncation order of the Josephson potential.

To arrive at an effective master equation, note that the system-bath interaction changes as well under the transformation to the effective frame
\begin{align}\label{eq:HsbI_collapse}
    e^{\hat{S}(t)/i\hbar}\hH_{sB}e^{-\hat{S}(t)/i\hbar} \; \equiv \hbar \sum_j \hat{C}(\omega_j) e^{-i\omega_jt} \otimes \hat{B}(t), 
\end{align}
with $\hat{B}(t)$ the bath operator in the interaction picture with respect to the bath Hamiltonian, and $\hat{C}(\omega_j)$ a time-independent collapse operator (here, a polynomial of the creation and annihilation operators of the two normal modes $a$ and $b$).  Following the procedure in~\cite{petrescu_lifetime_2020} one can derive an effective master equation from \cref{eq:HsbI_collapse}
\begin{align}\label{eq:L_eff}
    \mathcal{L}_{\textit{eff}} (\hat{\rho}) = \frac{1}{i\hbar} [\hat{K}, \hat{\rho} ] + \sum_j \kappa(\omega_j) \mathcal{D}_{\hat{C}(\omega_j)} (\hat{\rho}),
\end{align}
where $\omega_j$ are a set of transition frequencies to be determined in perturbation theory below, and $\kappa(\omega_j)$ the bilateral power spectral density of the noise. The above master equation is valid in the limit where the coupling $\kappa_b$ is much smaller than the transition frequencies $\omega_j$ of the system~\cite{petrescu_lifetime_2020}, as in the standard treatment of the secular, Born, and Markov approximation master equation~\cite{breuer_dissipative_1997}. The derivation is presented in \ref{app:sec:eff-Lind}.

% is the bilateral power spectral density of the noise
% \begin{align}
%     \begin{split} \label{eq:sbb}
%         \kappa(\omega)= \int_{-\infty}^{\infty} d \tau e^{-i \omega \tau} \operatorname{Tr}[ \hat{B}(\tau) \hat{B}(0) \hat{\rho}_B^0],
%     \end{split}
% \end{align}
% with $\hat{\rho}_B^0$ is the steady state density matrix of the bath~\cite{breuer_dissipative_1997}. We have neglected the Lamb shift in this derivation, as this is typically absorbed in a redefinition of the normal-mode frequencies $\omega_{a,b}$. The above master equation is valid in the limit where the coupling $\kappa_b$ is much smaller than the transition frequencies $\omega_j$ of the system~\cite{petrescu_lifetime_2020}, as in the standard treatment of the secular, Born, and Markov approximation master equation~\cite{breuer_dissipative_1997}.

To relate this to the familiar treatment of the master equation, if the system were undriven ($\epsilon_p, \epsilon_d= 0$), the system's Hamiltonian \cref{eq:Hs_g_ology} would be time-independent and diagonal in the tensor-product Fock space corresponding to the two normal modes $a$ and $b$, leading to $e^{\hat{S}(t)/i\hbar} = I$ and $\sum_j \hat{C}(\omega_j) e^{-i\omega_jt} = \hat{y}_b(t) + u\hat{y}_a(t) $ that yields the dissipators $\kappa_{a} \mathcal{D}_{\hat{a}}$ and $\kappa_{b} \mathcal{D}_{\hat{b}}$ in the master equation, with $\kappa_b = \kappa(\omega_d) = \kappa(\omega_b)$ and $\kappa_a = \kappa\left(\frac{\omega_d+\omega_p}{2}\right) = \kappa(\omega_a)$, as dictated by the second resonance matching condition \cref{eq:res-cond}.

\subsection{Results}\label{sec:eff-mod:results}
\begin{figure*}[t!]
    \centering
    \includegraphics[width=\textwidth]{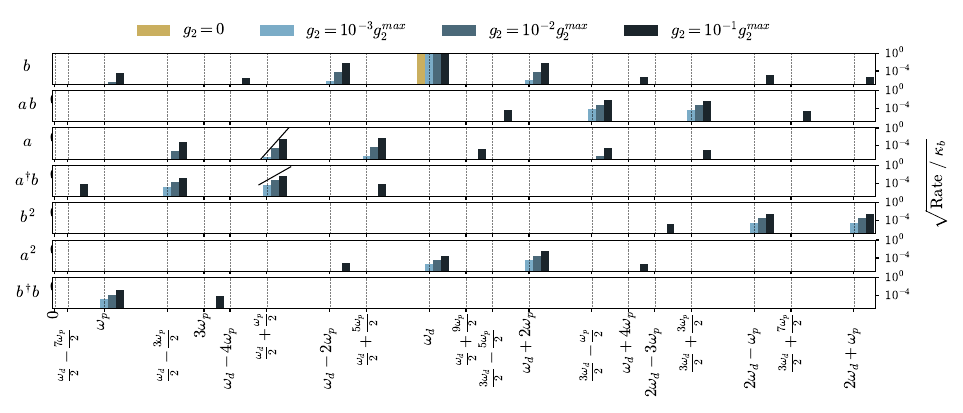}
    \caption{Analysis of drive-induced collapse operators in $O(\lambda^8)$ SWPT: Absolute value of the prefactor of the monomial written on the $y$-axis in the collapse operator $C(\omega_i)$ of the effective master equation corresponding to Liouvillian \cref{eq:L_eff}, whose frequency $\omega_i$ is given on the $x$-axis. The two solid  lines indicate the scaling in $g_2$ or $g_2^2$ of the prefactor, as discussed in the main text. The pump power is defined with the magnitude of the leading parametrically activated term $g_2$ with respect to its maximum value $g_{2}^\textit{max}/2\pi=50.8\text{MHz}$ from \cref{eq:gs_RWA}. The $\ha$ coupling features a second-order dependence on pump power. We used an experimental parameter set
    $\omega_a/2\pi = 4\;\mathrm{ GHz},
    \omega_b/2\pi = 7.05\;\mathrm{ GHz}, 
    \varphi_a = 0.11, 
    \varphi_b = 0.2, 
    E_J/h = 37\;\mathrm{ GHz}, 
    E_L/h = 62.4\;\mathrm{ GHz}$ and $\epsilon_d = 5 g_2$ corresponding to a cat state with $|\alpha|=\sqrt{5}$ [\cref{eq:H2pho}]. The $x$-axis positions of the bars are set by the frequencies of the charge drive and the flux pump $\omega_d/2\pi = 7.05\;\mathrm{ GHz}$ and $\omega_p/2\pi=0.95\;\mathrm{ GHz}$. The minimum of the $y$-axis is $10^{-7}$. For simplicity, we set $u=0$. We find that for $\epsilon_d = 0$ the difference of the absolute value of the prefactor of the monomials with the presented ones are smaller than $10^{-4}$, as we detail in~\cref{app:extra_collapse:ed}}
    \label{fig:Collapse_monom}
\end{figure*}

A computer-assisted calculation (see \cref{app:trunc}) results in around 2000 terms for the sixth-order expansion of the system-bath coupling Hamiltonian  \cref{eq:HsbI_collapse}. The sixth order in $\lambda$  turned out to be the lowest order at which we could observe effective single-photon losses induced by the pump onto mode $a$. The effective Hamiltonian reads
\begin{align}
    \begin{split}\label{eq:Heff_6}
        \hat{K}/\hbar =& (\delta + \delta^\prime) \ha^\dagger \ha + (\Delta + \Delta^\prime) \hb^\dagger \hb \\
        & - i g_2(\ha^2-\alpha^2) \hb^\dagger  +\text{h.c.}\\
        & + g_2^a (\ha^\dagger\ha) \ha^2\hb^\dagger  +\text{h.c.} \\
        & + g_2^b \ha^2 \hb^\dagger (\hb^\dagger \hb)  +\text{h.c.} \\
        &+ O(E_J\lambda^7).
    \end{split}
\end{align}
We recover to third order in $\lambda$ on the second row of \cref{eq:Heff_6} a flux-pump amplitude-dependent 2-1 photon interaction in \cref{eq:H2pho}, alongside a number of other coupling constants allowed up to sixth order in $\lambda$. In terms of the g-ology of \cref{eq:gs}, the coupling constants in the effective Hamiltonian \cref{eq:Heff_6} read
\begin{align}\label{eq:gs_RWA}
\begin{split}
     -i g_2=& \ 3 r \mathrm{g}_{3,1} = -i {E}_J J_1(\epsilon_p) \varphi_a^2\varphi_b,\\
     g_2^a =&\ 20 r \mathrm{g}_{5,1},\\
     g_2^b =&\ 30 r^3 \mathrm{g}_{5,1},\\
     \alpha^2=&\  -i\dfrac{\epsilon_d}{g_2},\\ 
     \delta^\prime =& 12 \mathrm{g}_{1,1} \mathrm{g}_{3,1} \left(\frac{r^{2}}{\omega_{d} + \omega_{p}} + \frac{2}{\omega_{d} + 3 \omega_{p}} + \frac{r^{2} + 2}{\omega_{d} - \omega_{p}}\right),\\
    \Delta^\prime =& 12 \mathrm{g}_{1,1} \mathrm{g}_{3,1} r^{2} \left(\frac{r^{2}}{\omega_{d} + \omega_{p}} + \frac{2}{\omega_{d} + 3 \omega_{p}} + \frac{r^{2} + 2}{\omega_{d} - \omega_{p}}\right).
\end{split}
\end{align}
The first equation above predicts a saturation of $g_2$ from this perturbation theory, with a maximum at $\epsilon_p \simeq 0.6 \pi$ that will be denoted by $g_2^{max}$. For the parameters in the caption of \cref{fig:Collapse_monom}, $g_2^\textit{max}/2\pi = 50.8\text{ MHz}$. In the rest of this work, we use $g_2/g_2^{max}$ to give a scale for the strength of the non-linear perturbation. Note that the charge drive on the buffer mode is such that $|\epsilon_d|= g_2 |\alpha|^2 = O(E_J\lambda^4)$, for moderate cat sizes. We have checked that the off-resonant terms coming from this charge drive have a negligible impact on the system-bath coupling (see \cref{app:extra_collapse}).

As mentioned above in our discussion of \cref{eq:res-cond}, we need to adjust the drive and pump frequencies to match the ac Stark shift of the system resonances. To this end, we collect the quadratic terms in the effective Hamiltonian $\hat{K}$ of \cref{eq:Heff_6}, and we choose the drive frequencies $\omega_p,\ \omega_d$ to cancel these terms
\begin{align}
\begin{split}\label{eq:pump_match}
    \omega_a - \frac{\omega_d + \omega_p }{2} + \delta^\prime(\omega_d,\omega_p) = 0,\\
    \omega_b - \omega_d + \Delta^\prime(\omega_d,\omega_p) = 0.
\end{split}
\end{align}
These can be recast as polynomial equations in $\omega_d$ and $\omega_p$, whose solution gives the drive frequencies that satisfy the resonance condition \cref{eq:res-cond}.

In \cref{fig:Collapse_monom}, we represent the terms of the effective system-bath coupling calculated to the seventh order in $\lambda$, to validate the sixth-order result. For a collapse operator identified by its frequency $\hat{C}(\omega_j)$ [see \cref{eq:HsbI_collapse}], we plot the leading prefactors in absolute value for the monomials ($\ha^{\dagger m}\ha^n \hb^{\dagger p} \hb^q$ with $m,n,p,q$ non-negative integers) appearing in $\hat{C}(\omega_j)$. At zero pump power, we find a single contribution of the form $\hb$ at frequency $\omega_d = \tilde{\omega}_b$, as expected. When increasing the pump power, we find parametrically activated dissipation mechanisms.  We identify the leading parity-breaking collapse operators to be $\ha\hb$, $\ha$, $\ha^\dagger \hb$. The prefactors corresponding to monomial $\ha$ are of second order in the pump power, as indicated by two solid lines showing linear and quadratic dependence on $g_2/g_2^\textit{max}$. We set $u=0$, and discuss the off-resonant dressing of the mode $a$ separately in \cref{app:extra_collapse}. A typical value of the normal mode hybridization is $u \simeq 0.06$.

When stabilizing cat states, the buffer mode remains close to the vacuum state while the $\ha$ mode is in a cat state~\cite{berdou_one_2023}. Therefore, we expect the $\ha$ coupling to the bath to directly affect the phase-flip rate of the stabilized cat states and the $\ha^\dagger \hb$ and $\ha \hb$ to have little impact on a stabilized cat state. However, these collapse operators can also result in phase-flips of a cat qubit when the buffer mode has a non-zero population while operations are performed on the system~\cite{gautier_designing_2023}. For instance, this occurs during the transient when preparing a cat state from the vacuum.

The effect of a non-zero temperature bath can be furthermore estimated from \cref{fig:Collapse_monom}. The sum of the collapse operators \cref{eq:HsbI_collapse} is a hermitian operator, and thus for every $\hat{C}(\omega)$ we have a hermitian conjugate $\hat{C}(-\omega)=\hat{C}^\dagger(\omega)$ which correspond to the complementary conversion process which is energetically forbidden. For a non-zero temperature bath, thermal photons can bring the necessary energy to overcome this barrier, and the obtained rate obeys the detailed balance $\kappa(\omega) = e^{\beta\hbar\omega}\kappa(-\omega)$ ~\cite{breuer_dissipative_1997}. For a dilution refrigerator at $T=10\;\mathrm{mK}$ the peaks at $\omega_p\simeq 1\;\mathrm{GHz}$ have thermally activated complementary processes with an amplitude less than $10\%$ that of the direct process. Note that the lowest frequency peak in \cref{fig:Collapse_monom} at $\frac{\omega_d}{2}-\frac{7\omega_p}{2}\simeq 200\;\mathrm{MHz}$ will have a complementary process in $\ha \hb^\dagger$ with an amplitude of the order of $35\%$ of the direct process. The effect of a non-zero temperature thermal bath remains negligible when interested in processes in the $\text{GHz}$-range.

In summary, we have derived using SWPT an effective model for the pump-induced transitions of the circuit and have identified leading parity-breaking processes. This effective model is expected to be accurate in the regime where $\lambda \ll 1$. In particular, the quantity $g_2/g_2^{max}$ has to be small compared to unity, to an extent which will be quantified below. In the next section, we assess the validity of our analytical formulas \cref{eq:gs_RWA} with increasing $g_2/g_2^\textit{max}$ by comparing them to exact Floquet numerical simulations.

\section{Comparison to exact numerical simulations}\label{sec:strong-drive}
In this section, we keep in line with the assumptions used in the development of the effective master equation in \cref{sec:eff-mod} and we restrict the analysis to a weakly dissipative mode $b$. However, instead of using SWPT to obtain the eigenspectrum of the driven Hamiltonian, we use Floquet numerical simulations~\cite{grifoni_driven_1998}, which allows us to explore regimes of relatively larger $g_2/g_2^\textit{max}$. For computational simplicity, we set the buffer drive amplitude to $\epsilon_d = 0$. In practice, the drive amplitude $\epsilon_d$ is of the order of $g_2\alpha^2$, where the number of photons  $|\alpha|^2$ of the stabilized cat state ranges typically smaller than 10~\cite{reglade_quantum_2024, marquet_autoparametric_2024}. Therefore, it is small compared to the linear drive terms of the sine function $\propto 2{E}_J \epsilon_p \varphi_a$ in \cref{eq:Hs_g_ology}, and results only in small corrections to the rates, as discussed in \cref{app:extra_collapse}. Consequently, we expect the range of validity of the effective model with $\epsilon_d=0$ to be similar to the one where $\epsilon_d \neq 0$ and cat states are stabilized.  In the following, we compare transition rates from SWPT with those obtained from Floquet numerical simulations, as a function of the flux-pump amplitude $\epsilon_p$, parametrized as the ratio $g_2/g_2^\textit{max}$, with $g_{2}^\textit{max}/2\pi=50.8\text{MHz}$ defined in \cref{sec:eff-mod}.

\subsection{Rates in Floquet Theory}
To set up a numerical solution to the Floquet eigenproblem, after putting $\epsilon_d=0$ in \cref{eq:H-circuit_1} 
to retain only a periodic time dependence, we have
\begin{align}
    \label{eq:tot_hamiltonian_fl}
    \begin{split}
        \hH_{s} =& \hbar \omega_a \ha^\dagger \ha + \hbar \omega_b \hb^\dagger \hb\\
        & - 2 E_J \sin\left[\epsilon(t)\right] \sin[\varphi_a ( \ha + \ha^\dagger ) + \varphi_b ( \hb + \hb^\dagger )].
    \end{split}
\end{align}
As we have done above, we simplify the system-bath coupling and set $u=0$ such that the buffer-mode induced Purcell decay of the memory mode $a$ is ignored, and we will address the relatively smaller contribution of this term in \cref{sec:aToBath}. Then, we obtain transition rates in the Born-Markov approximation~\cite{bluemel_dynamical_1991} using QuTip~\cite{johansson_qutip_2013}. In this approximation, relaxation times should remain small with respect 
to the transition frequencies of the system. We define the zero-temperature Floquet-theory transition rate matrix,
\begin{align}\label{eq:gamma_fl}
    \begin{split}
        \Gamma_{i \to j}^{(F)} &= \sum_k |y_{ijk}|^2 \Theta(\Delta_{ijk}) J(\Delta_{ijk}),\\
        y_{ijk} &= \frac{\omega_p}{2\pi} \int_0^{\sfrac{2\pi}{\omega_p}} dt \bra{\phi^{(F)}_i(t)} \hat{y}_b \ket{\phi^{(F)}_j(t)} e^{-i k \omega_p t} ,
    \end{split}
\end{align}
where the system is coupled to the bath through the operator $\hat{y}_b = -i(\hb-\hb^\dagger)$. Here, $\Gamma_{i \to j}^{(F)}$ is the transition rate from Floquet mode $|\phi_i^{(F)}(t)\rangle$ to $|\phi_j^{(F)}(t)\rangle$, $\Theta$ is the Heaviside step function since we assume the bath to be at zero temperature, and $\Delta_{ijk} =  \epsilon_i-\epsilon_j + k\omega_p$ is the transition frequency where $\epsilon_i$ is the $i$-th quasi-energy of the Floquet problem~\cite{grifoni_driven_1998}. Moreover, $J$ is the bath spectral function~\cite{bluemel_dynamical_1991}, assumed to be flat, \ie $J(\omega) = \kappa_b \Theta(\omega)$. We make this choice to simply evaluate the relative contributions of various monomials resulting from the Josephson nonlinearity without regard to the frequency dependence of the bath spectrum. The addition of filters, effectively modifying the bath spectral function, is discussed in \cref{sec:mitigation}.

Regarding convergence of the numerical solution to the Floquet eigenproblem, we choose the truncation of the Fock space dimension to be $20\times11$ for the modes $a$ and $b$, respectively. The validity of this truncation is assessed by requiring that the average of the canonical commutators $\langle [\hat{c},\hat{c}^\dagger]\rangle$, for $\hat{c}=\ha,\hb$,  for all relevant states, to be close to unity below $10^{-8}$. 

\subsection{Correspondence to SWPT and state tracking}
We now need to compare the rates obtained from Floquet numerical simulations \cref{eq:gamma_fl} to the effective model in \cref{sec:eff-mod} with the following definition. The Floquet eigenstates of the system can be expressed in terms of the Floquet modes~\cite{grifoni_driven_1998} and are directly related to the eigenvectors of the effective Hamiltonian $\hat{K}$ of \cref{eq:Heff_6}. In SWPT, the approximation to the $j^\textit{th}$ Floquet state $e^{-i\epsilon_j t}|\phi_j^{(F)}(t)\rangle$ is 
\begin{align} \label{eq:FlSWPT}
\ket{\psi_j^\text{SWPT}(t)} = e^{-i\hat{H}_0 t/\hbar}e^{-\hat{S}(t)/i\hbar} e^{- i\hat{K}t/\hbar} \ket{\psi_j},
\end{align} 
where $\ket{\psi_j}$ is an eigenvector of the time-independent Hamiltonian $\hat{K}$. Using the expression of the effective system-bath coupling, \cref{eq:HsbI_collapse}, we can express the transition rates between eigenstates using Fermi's Golden rule,
\begin{align}\label{eq:gamma_rwa}
    \begin{split}
        \Gamma_{i \to j}^\text{(SWPT)} &= \sum_\omega \kappa(\omega) \left| \bra{\psi_j}\hat{C}(\omega)\ket{\psi_i} \right|^2.
    \end{split}
\end{align}
The eigenvectors $\ket{\psi_i}$ are obtained via a numerical diagonalization of the Hamiltonian $\hat{K}$ \cref{eq:Heff_6}, expressed in the Fock basis and truncated. In the following, we need to compare the SWPT rate in \cref{eq:gamma_rwa} to the one obtained from Floquet theory in \cref{eq:gamma_fl}.

To appropriately associate transition rates to states in the spectrum, when incrementally increasing the pump power, we track Floquet modes using the maximum overlap with the Floquet modes obtained at the previous pump power. At zero pump power, the Floquet modes coincide with the eigenstates of the zero-pump Hamiltonian, which are photon-number $\hat{N}_{a,b}$ eigenstates. Moreover, we numerically find the ac Stark shifted frequency matching condition, by sweeping pump frequency at each pump power~\cite{petrescu_accurate_2023,xiao_diagrammatic_2023}. We use the condition \cref{eq:res-cond} as a starting point for the frequency sweep.

Note that, in order to compare rates obtained from \cref{eq:gamma_rwa} or from \cref{eq:gamma_fl}, we need to appropriately map states between the two methods.  \Cref{eq:FlSWPT} implies that at $t=0$ eigenvectors of $\hat{K}$ almost coincide~\cite{garcia-mata_effective_2024} with the Floquet modes $\vert \phi_i^{(F)}(0) \rangle = e^{-\hat{S}(0)/i\hbar}\ket{\psi_i} \simeq \ket{\psi_i}$. By looking for the largest overlap between Floquet modes and eigenvectors of the effective Hamiltonian $\hat{K}$, we can establish the required one-to-one correspondence.

\subsection{Defining rates in the presence of hybridization}
\begin{figure}[t!]
    \centering
    \includegraphics[width=\linewidth]{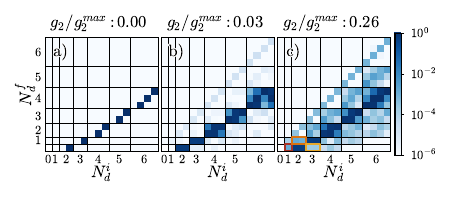}
    \caption{Transition rate matrices $\Gamma_{i \to f }^{(F)}/ \kappa_b$  in the Born-Markov approximation \cref{eq:gamma_fl} versus initial and final state for various pump amplitudes. The Floquet eigenstates are sorted by their mean-value of $\hat{N}_d$,\cref{eq:Nd}, on the x and y-axis. a) At zero pump power, the dissipation of the mode $b$ generates transitions that change excitation number $\hat{N}_d$ by $\pm 2$. The photon-number parity of mode $a$ is conserved, since transitions happen only between $N_d$ and $N_d-2$ sectors. b) Increasing the pump power strongly hybridizes the modes within a given $N_d$ sector but approximately conserves photon-number parity of mode $a$. c) At large pump powers the parity-breaking transitions, such as those connecting $\hat{N}_d$ and $\hat{N}_d\pm1$, start to be non-negligible. 
    We further analyze the sectors highlighted in c) in \cref{fig:G_a-0_eps}. Parameters as in \cref{fig:Collapse_monom}.}
    \label{fig:Fl-tran-matrix}
\end{figure}

\Cref{fig:Fl-tran-matrix} shows the transition matrix $\Gamma^{(F)}$ for various $g_2/g_2^{max}$. To classify these transition rates, we refer back to the lowest-order effective Hamiltonian \cref{eq:H2pho} at vanishing buffer drive $\epsilon_d=0$
\begin{align}\label{eq:H_eff_ed0}
    \hat{K}^{\epsilon_d=0} &= g_2 (\ha^2 \hb^\dagger +\text{h.c.} ).
\end{align}
This Hamiltonian, which neglects all corrections from off-resonant processes calculated in the previous \cref{sec:eff-mod:eff-ME}, determines symmetry sectors between which we can unambiguously define transition rates. It conserves the photon-number parity of the mode $a$, and the dressed excitation number 
\begin{align}
    \hat{N}_d = \hat{N}_a + 2\hat{N}_b, \label{eq:Nd}
\end{align} 
where $\hat{N}_{a/b}$ is the photon number operator of the normal mode $a$ and $b$ respectively. 

In the undriven system, \cref{fig:Fl-tran-matrix}a), $\hat{N}_{a,b}$ are separately conserved by the lowest-order effective Hamiltonian \cref{eq:H_eff_ed0}, whereas the capacitive coupling of mode $b$ to the bath induce transitions that change $\hat{N}_d$ by $\pm 2$. At small pump powers, \cref{fig:Fl-tran-matrix}b), photon numbers $\hat{N}_{a,b}$ are no longer good quantum numbers. Reverting to a description in terms of the dressed excitation number $\hat{N}_d$, transitions changing the latter by $\pm 2$  dominate by orders of magnitude over other transitions, including those that change $\hat{N}_d$ by $\pm 1$ or $\pm 3$. This is no longer the case when further increasing the pump power,~\cref{fig:Fl-tran-matrix}c). 

\subsection{Comparison between SWPT and Floquet Theory. Circuit impedance}
\label{sec:Comparison}

\begin{figure*}[t]
    \centering
    \includegraphics[width =\linewidth]{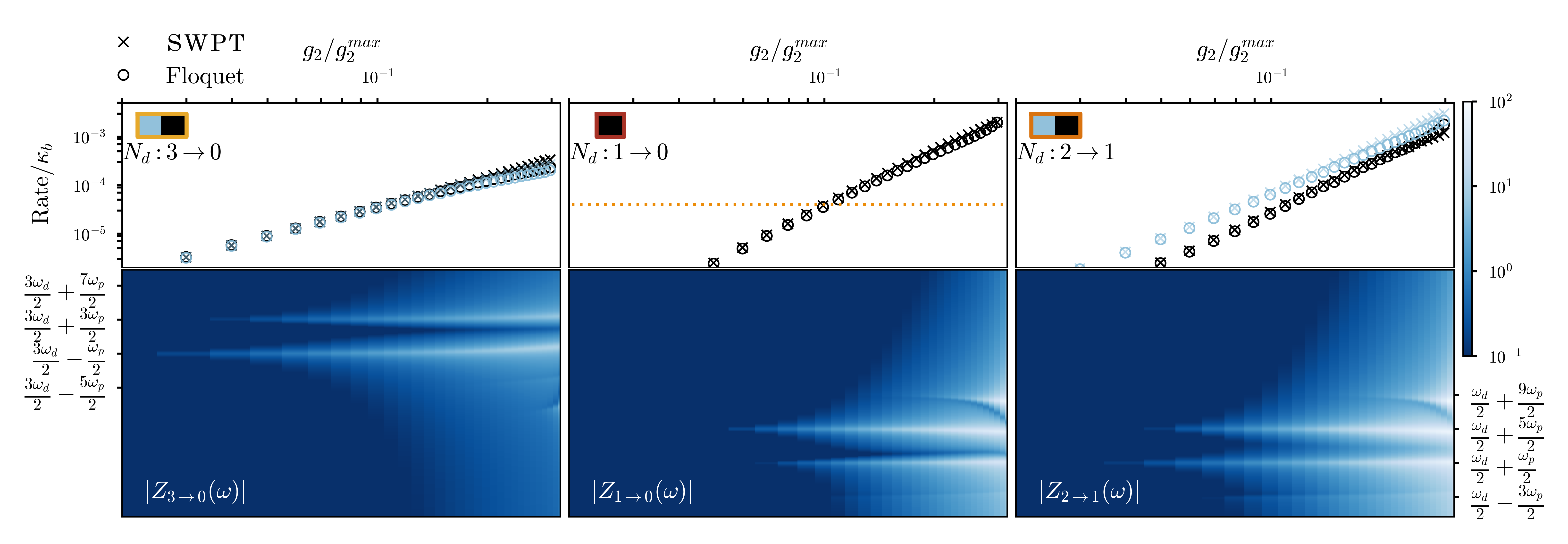}
    \caption{Transition rates corresponding to leading parity-breaking monomials identified in \cref{fig:Collapse_monom} $\ha\hb$, $\ha$ and $\ha^\dagger \hb$ from right to left. As explained in the text, each column corresponds to one or multiple transitions between a pair of $N_d$ sectors [colored rectangles in \cref{fig:Fl-tran-matrix}c)]. The top line compares the transition rates computed within Floquet-Markov theory~\cref{eq:gamma_fl} and the effective transition rates from \cref{eq:gamma_rwa} when increasing the pump power. The rates are computed between the tracked eigenstates of the system. The orange dotted line in the panel $N_d:1\to0$ corresponds to the threshold of the repetition code \cref{eq:tsep}. The panels on the bottom show the partial impedance associated with the transitions on the above panels (see \cref{app:abs_spect}). For comparison with the spectral features of \cref{fig:Collapse_monom}, the $y$-axis was labeled with the predicted frequencies of the collapse operators.
    }
    \label{fig:G_a-0_eps}
\end{figure*}

In view of the discussion in the previous subsection, the three effective dissipators giving the largest contributions to the effective system-bath coupling as derived in \cref{fig:Collapse_monom}, namely $\ha$, $\ha^\dagger\hb$, and $\ha\hb$, correspond to transitions that change the dressed excitation number as follows $N_d:1\to0$, $N_d:2\to1$, and $N_d:3\to0$ (highlighted in \cref{fig:Fl-tran-matrix}c). We check the agreement of the SWPT with Floquet numerical simulations by plotting the transition rates for all pairs of states  corresponding to the initial and final dressed excitation numbers as a function of $g_2/g_2^{max}$ in \cref{fig:G_a-0_eps}. We represent Floquet transition rates of the leading parity-breaking monomials identified in \cref{sec:eff-mod} along those obtained in the effective master equation in \cref{sec:eff-mod:eff-ME}. We report a relative error of less than $10\%$ at $g_2/g_2^\textit{max}=0.1$. For a more detailed analysis of the agreement between SWPT and Floquet numerical simulations, see \cref{app:agreement}. The red dotted line in the middle left panel of \cref{fig:G_a-0_eps} denotes the threshold of the repetition code for a cat-size $\alpha=\sqrt{8}$ [see \cref{sec:intro}].

The transitions described above lead to spectral peaks in the frequency-dependence of the response of the circuit.  In \cref{app:abs_spect} we derive a formula for the impedance of the Floquet system using a Kubo formula~\cite{kubo_statistical-mechanical_1957,pietikainen_observation_2017}. It is related to the reflection coefficient under a probe signal coming from the transmission line (see \cref{fig:GalvaCat}). We further define the partial impedance corresponding to transitions between Floquet modes $i$ and $j$
\begin{align}
\begin{split} \label{eq:Zij}
    Z_{i\to j}(\omega) \propto \sum_{k} & \left|y_{ijk}\right|^2 \left[\frac{1}{\Gamma^{(i,j)} + i(\omega-\Delta_{ijk})}\right.\\
    &\left.- \frac{1}{\Gamma^{(i,j)} + i(\omega + \Delta_{ijk})} \right],
\end{split}
\end{align}
where $\Gamma^{(i,j)}=  \sum_n \frac{\Gamma_{j \to n}^{(F)} + \Gamma_{i \to n}^{(F)} }{2} - \delta_{i,j} \Gamma_{i \to i}^{(F)}$ corresponds to the average of the inverse lifetime of the states $i$ and $j$. The linewidth of the mode $b$ is set to $\kappa_b/2\pi = 100\text{ MHz}$ which corresponds to a working point for the stabilization of cat states in the range $g_2/g_2^\textit{max} \geq 0.1$~\cite{marquet_autoparametric_2024}.

We plot the sum of the partial impedances \cref{eq:Zij} relating the sector $N_d^i$ to $N_d^f$, $Z_{N_d^i\to N_d^f}(\omega)~=~\sum_{i\in  N_d^i, f\in N_d^f} Z_{i\to f}(\omega)$, on the right-hand panels of \cref{fig:G_a-0_eps}. 
The peaks in the partial impedance should be compared to the frequencies of the collapse operator containing the corresponding monomials \cref{fig:Collapse_monom}. For instance, the effective 1-photon losses related to the monomial $\ha$ in \cref{fig:Collapse_monom} occur through the emission of photons in the bath at the frequencies $\frac{\tilde{\omega}_b}{2} -\frac{3\omega_p}{2} $, $\frac{\tilde{\omega}_b}{2} +\frac{\omega_p}{2} $ and $\frac{\tilde{\omega}_b}{2} +\frac{5\omega_p}{2} $, which is consistent with the features of $|Z_{1\to 0}|$. Similarly, the spectral features of $|Z_{2\to1}|$ and $|Z_{3\to0}|$ correspond to the frequencies predicted  in \cref{fig:Collapse_monom}.
 
To summarize, the spectral analysis of the collapse operators involved in the parity-breaking monomials in \cref{fig:Collapse_monom} is in good agreement with the spectral features of the corresponding partial impedance \cref{fig:G_a-0_eps}. We have shown that the effective model derived in \cref{sec:eff-mod} gives precise estimates of the spectral properties in a range where $\lambda \simeq 0.1$ and for a pump amplitude up to $g_2/g_2^{\textit{max}}\simeq 0.2$. 
The analysis of the frequency response allows us to set constraints on, for example, Purcell filters on the buffer mode, a topic to which we turn in the next section.

\section{Mitigation techniques}
\label{sec:mitigation}
\subsection{Canceling the linear drive term}
As explained at the end of \cref{sec:eff-mod}, in the experimentally relevant regime, the main mechanism that will induce phase-flip errors in stabilized cat qubits is the $\ha$-like dissipation. From the effective analytical model presented in \cref{sec:eff-mod}, \cref{fig:Collapse_monom}, we find that the dominant contributions to the collapse operator $\ha$ are proportional to $\mathrm{g}_{1,1}$ in \cref{eq:gs}. The Hamiltonian term proportional to the coupling $\mathrm{g}_{1,1}$ corresponds to linear drive terms at frequencies $\pm \omega_p$ on both normal mode $a$ and normal mode $b$. This drive term can be removed by performing a linear displacement transformation on the modes, such as $\hc \to \hc + \xi_c^{(-)} e^{-i\omega_p t} + \xi_c^{(+)} e^{i\omega_p t}$ where $\hc = \ha,~\hb$, and where the field amplitudes $\xi_c^{(\pm)} \propto \mathrm{g}_{1,1}$.  Under this transformation, the third-order term in \cref{eq:Hs_g_ology}  resulting in off-resonant terms of the form  $g_2e^{-2i\omega_p t} \ha^2 \hb^\dagger +\text{h.c.} $ becomes $g_2\xi \ha \hb^\dagger e^{-i(\omega_a+2\omega_p-\omega_b)t} + \text{h.c.} $, corresponding to a process where one photon in the mode $a$ and two pump photons convert to an off-resonant photon in the mode $b$ at $\omega_a+2\omega_p$. This process results in an effective decay of the mode $a$ at frequency $\omega_a+2\omega_p = \frac{\omega_d+5\omega_p}{2}$. 

Therefore, to reduce the effective single-photon loss rate on the mode $a$, we can leverage the ATS flux degrees of freedom [\cref{fig:GalvaCat}b)] to cancel the linear drive term $\mathrm{g}_{1,1}$. 
The ATS circuit is threaded with two external fluxes $\Phi_L $ and $\Phi_R$, threading respectively the left and right loops. The nonlinearity is driven through $\Phi_\Sigma = (\Phi_L + \Phi_R)/2$, making  $g_2$ only a function of $\Phi_\Sigma$. Applying the rules on the assignment of time-dependent fluxes~\cite{you_circuit_2019} (see \cref{sec:irrot}), the differential flux $\Phi_\Delta = (\Phi_L - \Phi_R)/2$ couples to the central inductance only, resulting in a linear drive term, whereas the symmetric flux $\Phi_\Sigma$ enters both the linear and the nonlinear terms
\begin{align}
\begin{split}\label{eq:fluxcanceling_H}
    \hH_s(t) =& \hbar \omega_a \ha^\dagger \ha + \hbar \omega_b \hb^\dagger \hb \\
    & -E_L^{\textit{eff}}(\epsilon_p,\eta_p)\left[ \frac{e^{i\omega_p t}}{2i} + \text{c.c.}\right] \left(\varphi_a \hat{x}_a + \varphi_b \hat{x}_b \right)\\
    & - 2 E_J \sin\left[\epsilon(t)\right] \sin\left(\varphi_a \hat{x}_a + \varphi_b \hat{x}_b \right),
\end{split}
\end{align}
with $\phi_\Sigma = \frac{\pi}{2} + \epsilon_p\sin(\omega_p t)$ and $\phi_\Delta = \phi_\Delta^0 + \eta_p \sin(\omega_p t)$ (see \cref{app:Norm} for derivation and definitions). We can recover the effective model of \cref{eq:Hs_g_ology} by redefining $\mathrm{g}_{1,1}$,
\begin{align}\label{eq:cancel_cond}
    \mathrm{g}_{1,1} & =2i{E}_J J_{1}(\epsilon_p) \varphi_a - \frac{E_L^{\textit{eff}}(\epsilon_p,\eta_p)}{2i} \varphi_a.
\end{align}
The cancellation of linear terms in the Hamiltonian requires, to lowest order, $\mathrm{g}_{1,1} = 0$. Since the inductive energy is linear, we write $E_L^\textit{eff}(\epsilon_p,\eta_p) = E_{L\epsilon}^\textit{eff}\epsilon_p + E_{L\eta}^\textit{eff}\eta_p$, with $E_{L\eta}^\textit{eff}$ and $E_{L\epsilon}^\textit{eff}$ set by circuit parameters (see \cref{app:Norm}). For $\epsilon_p\ll \pi$, this sets the ratio $\frac{\eta_p}{\epsilon_p} = \frac{-2E_J-E_{L\epsilon}^\textit{eff}}{E_{L\eta}^\textit{eff}}$. As long as $E_{L\eta}^\textit{eff} \neq 0 $ there will always exist a ratio that cancels $\mathrm{g}_{1,1}$. A corresponding experimental signature is the cancellation of the ac Stark shift [see \cref{eq:Heff_6}] to the lowest order in $\lambda$.

\begin{figure}[t]
    \centering
    \includegraphics[width = \linewidth]{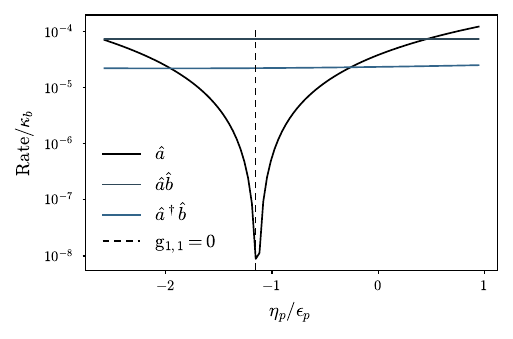}
    \caption{Leading effective rates identified in \cref{fig:Collapse_monom} versus $\eta_p/\epsilon_p$. Sweeping the ratio $\eta_p/\epsilon_p$ is equivalent to sweeping $\mathrm{g}_{1,1}$. At the cancellation point the effective rate in $
    \ha$ is significantly reduced. This is not true for all the effective rates identified in  \cref{fig:Collapse_monom}. The dashed line represents the cancellation condition \cref{eq:cancel_cond} in the limit $\epsilon_p \ll \pi$. The parameters are $E_{L\eta}^\textit{eff}/h = 62.4~\mathrm{GHz}$ and $E_{L\epsilon}^\textit{eff}=0$}
    \label{fig:flux_cancel}
\end{figure}

\Cref{fig:flux_cancel} shows the single-photon loss rate of the mode $a$, as determined in the effective master equation, as a function of the ratio $\eta_p/\epsilon_p$. We find that for $\mathrm{g}_{1,1}=0$ the rate is significantly reduced, but not canceled since higher-order effects beyond the $g_{1,1}$ terms are still causing effective single-photon losses. Nonetheless, other spurious decay processes identified in \cref{fig:Collapse_monom} are not suppressed.

We have shown that the analytical expressions derived from the effective model in \cref{sec:eff-mod} allow us to identify the origin of a given spurious process and then leverage the flux degrees of freedom of the ATS to mitigate this source of decoherence. In particular, we have shown that we can effectively neutralize single-photon losses by canceling linear terms in the system Hamiltonian. We emphasize that essential to this analysis was a correct treatment of time-dependent fluxes through the two loops of the ATS (\cref{app:Norm}).  With analytical expressions for the effective Liouvillian at our disposal, we can moreover sweep parameters with a small numerical cost. In the next section, we analyze the dependence of the collapse operators on the frequency choices and extract the parameter regime where the parity-breaking transitions can be mitigated. 

\subsection{Mitigation through design constraints}

\begin{figure*}[t!]
    \centering
    \includegraphics[width=\linewidth]{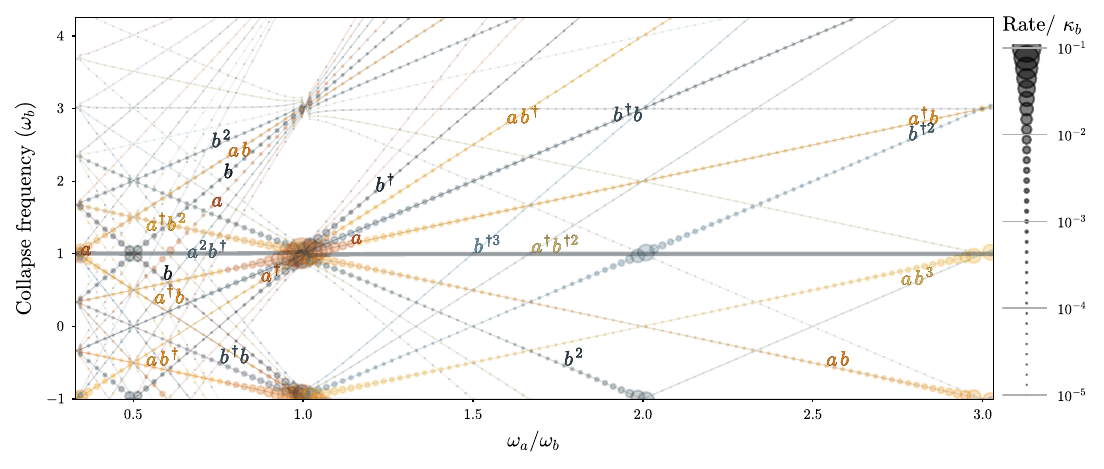}
    \caption{Collapse operators and associated rates as a function of the undressed memory mode frequency to $7^\textit{th}$ order in $\lambda$. Given a value of $\omega_a/\omega_b$ ($x$-axis), collapse operators appear at various collapse frequencies ($y$-axis).  
    The color is given by the monomial with the largest prefactor in the given collapse operator. Shades of blue denote parity-conserving monomials, while shades of red denote parity-breaking monomials. Annotations specify the concerned monomial and the associated rate corresponds to the radius of the dots. For dots below a threshold ($20\%$ of the largest dot at a given abscissa), we do not label the process. In this plot $ g_2 / g_2^{max} =0.1$ and other parameters as in \cref{fig:Collapse_monom}. The drive $\eta_p$ is chosen as in \cref{eq:H-circuit_1}.}
    \label{fig:collisions}
\end{figure*}

Another possible way to mitigate parity-breaking transitions is to change the parameter regime. Since the amplitude of effective dissipation rates is inversely proportional to linear combinations of the normal mode frequencies $\omega_a, \omega_b$, or equivalently $\omega_p, \omega_d$, we expect a strong dependence of the dominant system-bath coupling with respect to frequency. By changing the unit of energy to be $\hbar \omega_b$, we find that a sweep of $\omega_a /\omega_b$ is sufficient to explore all possible frequency regimes. In general, SWPT does not allow sweeping the frequency since the expansion is only valid far from the resonance (\ie near resonances small frequency detunings in the denominators make the expansion divergent). We address this issue by avoiding the frequencies at which a detuning smaller than the linewidth of the mode $b$, $\kappa_b\simeq 100~\mathrm{MHz}$ appears. In practice, the spacing between the frequencies displayed in \cref{fig:collisions} is large compared to $\kappa_b$, therefore only isolated points were removed.

In \cref{fig:collisions} the collapse frequencies are plotted versus the static normal-mode frequency of the mode $a$. At each such $\omega_a$ and in each collapse operator associated with a collapse frequency on the $y$-axis, we identify the monomial with the largest numerical prefactor and encode the value of this prefactor in the radius of the corresponding circle. The colors identify the dominant monomials uniquely with shades of red for parity-breaking monomials and shades of blue for parity-preserving monomials (for clarity only the leading monomials are labeled). 

For $\omega_a > \omega_b$, spurious induced decays get sparser and wider apart in frequency. Therefore, as the transmission line typically has a finite bandwidth, we can predict that high-frequency memories are less affected by spurious decay processes. Moreover, \cref{fig:collisions} gives constraints on the bandwidth of a Purcell-like filter to ensure that the system is not limited by off-resonant spurious decays \cite{putterman_preserving_2024, chamberland_building_2022}.  

Lines crossing the collapse frequency of the buffer correspond to accidental resonances in the original time-dependent Hamiltonian. An accidental resonance occurs when the two frequencies that characterize the resonance conditions \cref{eq:res-cond} satisfy $\omega_d p = q \omega_p$, with two integers $p,q$. 
%As an example, we show how a spurious resonance, reported in the recent experiment on ATS-based dissipative cat qubit~\cite{putterman_preserving_2024} with mode frequencies $\omega_a/\omega_b = 1.78$, shows up in our analysis. The identified resonance term was of the form 
For example, close to $\omega_a/\omega_b = 1.7$, we report a spurious decay mechanism of the form $\ha^\dagger \hb^{2\dagger}$, corresponding to an accidental resonance $3\omega_p = 5\omega_d$. This decay channel originates from a Hamiltonian term $\ha^\dagger \hb^{3\dagger}$.
This resonance was pointed out and studied in the recent experiment on ATS-based dissipative cat qubit~\cite{putterman_preserving_2024} with mode frequencies $\omega_a/\omega_b = 1.78$. 

To summarize, in this section, we have explored two methods to mitigate spurious decays induced by the pump. One method consists of reducing the number of circulating pump photons in the modes to suppress the dominant spurious decay process identified above. Secondly, we give a precise classification of the transitions, that can be used to identify optimal parameter regimes for the memory and buffer mode frequencies. 

\section{Conclusion}
\label{sec:conclusion}
We have derived an effective model for a dissipative cat qubit circuit using time-dependent Schrieffer-Wolff perturbation theory. We have seen how the parametric pumping scheme effectively modifies the system-bath coupling and classified the various contributions as a function of transition frequency and collapse operator. We assessed the validity of the model by comparing it to exact Floquet numerical simulations in the limit of weak system-bath coupling. 

In the case of a dissipative cat qubit stabilized by an ATS-based circuit, our study reveals that, in general, the ratio $\kappa_1/\kappa_2 $ eventually increases as a function of $\kappa_2$ for experimentally relevant circuit parameters, which degrades the noise bias required for quantum error correction. We show that the processes responsible for this increase can be mitigated by leveraging a careful treatment of the time-dependent external fluxes on the ATS to reduce the number of circulating pump photons. Finally, the analytical results from perturbation theory can be fed into the design of the system frequencies and the filtering of the transmission lines.

\section*{Acknowledgment}
We acknowledge useful discussions with Nicolas Didier, Jérémie Guillaud, Sébastien Jezouin, Raphaël Lescanne, Paul Magnard, Anil Murani, and Felix Rautschke. Computational tasks were performed on the Inria CLEPS cluster. We also thank the whole Alice \& Bob and Inria Quantic teams for creating the environment to perform this work. This project has received funding from the European Research Council (ERC) under the European Union’s Horizon 2020 research and innovation program (grant agreement No. 884762). 

\appendix
\section{Circuit Hamiltonian}
\label{app:Norm}
\begin{figure}[H]
    \centering
    \includegraphics[width = 0.8\linewidth]{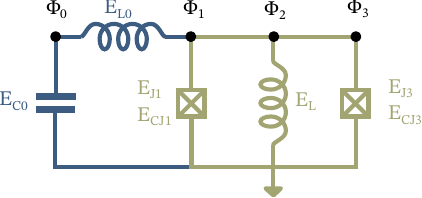}
    \caption{Diagram for the  quantization of the circuit of \cref{fig:GalvaCat}b) with node flux assignments.}
    \label{fig:GalvaCat_node}
\end{figure}

In this appendix, starting from the circuit diagram in \cref{fig:GalvaCat}b), quantizing the circuit with the time-dependent flux drives~\cite{you_circuit_2019}, we obtain a 4-variable circuit Hamiltonian and two holonomic constraints~\cite{you_circuit_2019} set by the time-dependent fluxes that drive the two loops of the ATS. We derive the Hamiltonian of the ATS (the green part of the node-labeled circuit in \cref{fig:GalvaCat_node}) in \cref{sec:irrot}. We carry on with the derivation of the full Hamiltonian of the circuit in \cref{fig:GalvaCat_node} in \cref{sec:modred}, which yields the Hamiltonian \cref{eq:H-circuit_1} that is our starting point in the main text. We then prove in \cref{app:bathcoupl}, based on the derived Hamiltonian, that dissipation induced by coupling to flux degrees of freedom is negligible compared to dissipation coming from charge noise.

\subsection{Irrotational constraint and lab-frame quantization}
\label{sec:irrot}

We first express the Lagrangian of the ATS in terms of the node fluxes defined in \cref{fig:GalvaCat_node}
\newcommand{\hQ}{\hat{Q}}
\newcommand{\hPhi}{\hat{\Phi}}
\begin{align}
\begin{split}
        \mathcal{L}\left(\Phi_{1,2,3},\dot{\Phi}_{1,2,3} \right)= \frac{C_{J1}}{2}\dot\Phi_1^2 + \frac{C_{J3}}{2}\dot\Phi_3^2 + \frac{C_L}{2}\dot\Phi_2^2\\
    + E_{J1}\cos\left( \phi_1\right)+E_{J3}\cos\left(\phi_3\right)-\frac{\Phi_2^2}{2L}, \label{eq:CalL} 
    \end{split}
\end{align}
where the junction capacitances are $C_{Ji}$ and Josephson energies are $E_{Ji}$, for both junctions on node $i=1,3$ in \cref{fig:GalvaCat_node}, the superconducting phase differences across the ATS junctions and inductor are related to the corresponding node fluxes via $\phi_{1,2,3} = 2\pi \frac{\Phi_{1,2,3}}{\Phi_0}$ with $\Phi_0$ the superconducting flux quantum. We define $C_L$ the capacitive shunt associated to the central inductor $L$, which satisfies $C_L \ll C_{Ji}$. Since there is only one superconducting node independent from ground, there are two time-dependent holonomic constraints imposed by flux quantization~\cite{you_circuit_2019}, which read
\begin{align}
    \begin{split}
        \Phi_l(t)=& \Phi_2 -\Phi_1, \\
        \Phi_r(t)=&\Phi_3 -\Phi_2,  \label{eq:Holo}
    \end{split}
\end{align}
where $\Phi_l (t)$ and $\Phi_r(t)$ are the external fluxes threading the left and right loops respectively.

Following You \textit{et al.}~\cite{you_circuit_2019}, we need to incorporate the holonomic constraints \cref{eq:Holo} to express our classical Hamiltonian in terms of a single branch flux $\Tilde{\Phi}$, expressed without loss of generality as a linear combination of the original  branch fluxes
\begin{align}
    \tilde{\Phi}=m_1\Phi_1+m_2\Phi_2+m_3\Phi_3. \label{eq:TPhi}
\end{align}
More specifically, inverting the definition of the node flux $\tilde{\Phi}$ \cref{eq:TPhi} together with the two holonomic constraints \cref{eq:Holo}, we reexpress the Lagrangian $\mathcal{L}(\Phi_{1,2,3}, \dot{\Phi}_{1,2,3})$ as $\mathcal{L}(\tilde{\Phi}, \dot{\tilde{\Phi}})$, with explicit dependence on the control fluxes $\Phi_{r,l},\dot{\Phi}_{r,l}$. We can then evaluate the momentum conjugate to this flux $Q = \partial \mathcal{L}/\partial \dot{\tilde{\Phi}}$  and perform the Legendre transform to get the classical Hamiltonian of the system 
\begin{align}
\label{eq:classicalHATS}
    \begin{split}
        \mathcal{H}(m_1,m_2,m_3) =&  \frac{m_{123}^2 Q^2}{2 C_{J1J3L}}  +\frac{1}{C_{J1J3L}} Q \left(\dot\Phi_r\alpha_r + \dot\Phi_l \alpha_l \right) \\
        &-E_{J1}\cos\left(\frac{\Tilde{\phi}-m_3\phi_r-m_{23}\phi_l}{m_{123}}\right)\\
        &-E_{J3}\cos\left(\frac{\Tilde{\phi} +m_{12}\phi_r+m_{1}\phi_l}{m_{123}}\right)\\
        &+\frac{\left(\Tilde{\Phi}+m_1\Phi_l-m_3\Phi_r \right)^2/m_{123}^2}{2L}\\
        &+F(\dot\Phi_r,\dot\Phi_l,\Phi_r,\Phi_l),
    \end{split}\end{align}
with $m_{ij}=m_i+m_j$ and $m_{ijk}=m_1+m_2+m_3$ and similarly for $C_{ijk}$.  Moreover, we introduced two coefficients that parametrize the charge drive induced by the derivatives of the flux drives [second term of \cref{eq:classicalHATS}],
\begin{align}
    \begin{split}
    \label{eq:alpharl}
    \alpha_r = -m_3C_{J1}+m_{12}C_{J3}-m_3C_L,\\
    \alpha_l = -m_{23}C_{J1}+m_1C_{J3}+m_1C_L. 
    \end{split}
\end{align}
Finally, note that $F$ in \cref{eq:classicalHATS} is a function of external fluxes only, which is immaterial to the resulting quantum theory since it can be absorbed in a redefinition of the ground-state energy.

The final step is to apply the irrotational constraints~\cite{you_circuit_2019}, consisting of canceling the charge drive induced by the time derivatives of the flux drives. For this, we need
\begin{align}
\begin{split}
    \alpha_r = \alpha_l = 0,
\end{split}
\end{align}
which, via \cref{eq:alpharl}, imposes two conditions on the three parameters $m_{1,2,3}$. Noting that the value of $m_{123}$ can be absorbed by a canonical transformation on the pair of coordinates $Q, \tilde{\Phi}$, we set the remaining condition on the three parameters $m_{1,2,3}$ as
\begin{align}
    m_{123} = m_1 + m_2 + m_3 = 1.
\end{align}
Moreover, upon taking the $C_L/C_{Ji}\rightarrow 0$ limit,  $m_2\rightarrow0$,  and the node flux 2 decouples from $\Tilde{\Phi}$, to give 
\begin{align}
    \begin{split}
        \mathcal{H} = \frac{{Q}^2}{2C_{J1J3}} &+ \frac{1}{2L}\left(\Tilde{\Phi} + \frac{C_{J1}}{C_{J1J3}}\Phi_l - \frac{C_{J3}}{C_{J1J3}}\Phi_r \right)^2 \\
        &-E_{J1}\cos\left[\Tilde{\phi} - \frac{C_{J3}}{C_{J1J3}}\left(\phi_r+\phi_l \right)\right]\\
        &-E_{J3}\cos\left[\Tilde{\phi} + \frac{C_{J1}}{C_{J1J3}}\left(\phi_r+\phi_l \right)\right].\\
    \end{split}
\end{align}
In the case of symmetric Josephson junctions we set $C_{J1}=C_{J3}=C_J$ and $E_{J1} = E_{J3}=E_J$, to obtain
\begin{align}
    \begin{split}
    \mathcal{H} = \frac{{Q}^2}{4C_J} &+ \frac{1}{2L}\left(\Tilde{\Phi} - \Phi_\Delta \right)^2 \\
        &-2E_{J}\cos(\phi_\Sigma)\cos\left(\Tilde{\phi}\right),
    \end{split}
\end{align}
with $\Phi_\Delta = \frac{\Phi_r- \Phi_l}{2}$ and $\phi_\Sigma = \frac{\phi_l+\phi_r}{2}$.

\subsection{Model reduction and normal modes}
\label{sec:modred}
In this section, we derive the laboratory frame Hamiltonian of the full circuit \cref{fig:GalvaCat_node} (\cref{app:deriv_labframe}).
After performing a normal-mode transformation (\cref{app:deriv_normal}), 
we apply an additional unitary transformation (\cref{app:deriv_disp}) to obtain the Hamiltonian of \cref{sec:eff-mod}, which models most of the ATS-based cat-qubit circuits. We then discuss tolerance to asymmetries in the ATS and to dc flux miscalibration in \cref{app:deriv_asymm} and \cref{app:deriv_flux_tol}.

\subsubsection{Lab-frame Hamiltonian of the full circuit}
\label{app:deriv_labframe}
The Lagrangian for the galvanically coupled circuit in \cref{fig:GalvaCat_node}, is, in terms of fluxes $\mathbf{\Phi} = \left(\Phi_0,\Phi_1,\Phi_2,\Phi_3\right)^T$,
\begin{align}\label{app:eq:Lagrange_Galva}
    \mathcal{L} &= \dot{\mathbf{\Phi}}^T \mathbf{C} \dot{\mathbf{\Phi}} + \mathcal{I}\left(\mathbf{\Phi}\right),
\end{align}
with the same holonomic constraints \cref{eq:Holo}, and
\begin{align}
\begin{split}
    \mathcal{I}(\Phi)&= E_{L_0} \frac{(\phi_0 - \phi_1)^2}{2} + E_{L} \frac{\phi_2^2}{2} \\
    &-E_J\cos(\phi_1) -E_J\cos(\phi_3).\\
    \mathbf{C} &= \text{diag}\left[\frac{C_0}{2} ,\frac{C_{J1}}{2} ,\frac{C_L}{2} ,\frac{C_{J3}}{2} \right],
\end{split}
\end{align}
with $E_{L_0}, E_L$ the inductive energies defined in \cref{fig:GalvaCat_node}, $E_J$ the Josephson energy, $C_0, C_{J1}, C_{J3}, C_L$ the capacitances of the node $0$, the Josephson junctions and the central inductance [see \cref{fig:GalvaCat_node} ].

We define 
\begin{align}
    \begin{split}
        \tilde{\mathbf{\Phi}} &= \mathbf{M}_{2\times4} \mathbf{\Phi},\\
        \mathbf{\Phi}_\textit{ext} &= \mathbf{R}_{2\times 4} \mathbf{\Phi}. \label{eq:PhiTilde}
    \end{split}
\end{align}
We introduced two real-valued matrices $\mathbf{M}_{2\times4}$ and $\mathbf{R}_{2\times4}$, whose subscripts indicate their dimensions.  $\mathbf{M}_{2\times4}$ characterizes the linear combination from the circuit variables $\mathbf{\Phi}$ to the independent variables $\tilde{\mathbf{\Phi}}$, while $\mathbf{R}_{2\times 4}$ encodes the constraints that relate the external fluxes $\mathbf{\Phi}_\textit{ext}$ to the node fluxes $\mathbf{\Phi}$. 
 
As before, we need to solve the irrotational constraint equation for $\mathbf{M}_{2\times4}$. This can be written as follows,
\begin{align}
    \mathbf{R}_{2\times 4} \mathbf{C}^{-1}(\mathbf{M}_{2\times4})^T =  \mathbf{0}.
\end{align}

Taking the limit $C_L \to 0$, we find
\begin{align}
    \mathbf{M}_{2\times4} = \left(\begin{matrix}1 & 0 & 0 & 0 \\0 & \frac{c}{c+1} & 0 & \frac{1}{c+1} \end{matrix}\right),
\end{align}
so that, by \cref{eq:PhiTilde}, $\tilde{\Phi}_0 = \Phi_0$ and $\tilde{\Phi}_2 = \Phi_\text{ATS} = \frac{c\Phi_1+\Phi_3}{(c+1)}$. We introduced $c=\frac{C_{J1}}{C_{J3}}$ the ratio of the capacitances. After performing the Legendre transformation and quantizing the fields, we find the Hamiltonian operator 
\begin{align}
    \begin{split}
    \hH & = \frac{ \hQ_{\text{ATS}}^{2}}{2(C_{J1}+C_{J3})} + \frac{ \hQ_{0}^{2}}{2C_{0}} + E_{L_0} \frac{ \left(\hat{\phi}_\text{ATS} - \hphi_{0} - \frac{2}{c+1}\phi_{\Sigma}\right)^{2}}{2}\\
    &+ E_L \frac{ \left(\hat{\phi}_\text{ATS} - \phi_{\Delta} + \frac{c-1}{c+1}\phi_\Sigma\right)^{2}}{2}\\
    & - E_{J1} \cos{\left(\hat{\phi}_\text{ATS} - \frac{2}{c+1}\phi_{\Sigma} \right)}\\
    &- E_{J3} \cos{\left(\hat{\phi}_\text{ATS} + \frac{2c}{c+1}\phi_{\Sigma} \right)},
    \end{split}
\end{align}
with $\hQ_i,\hPhi_i, \hat{\phi}_i$ charge, flux and phase operators respectively. We introduce the dimensionless operators
\begin{align}
    \begin{split}
        \hat{\phi}_0 &= \left(\frac{2E_{C_0}}{E_{L_{0}}}\right)^{1/4} \hat{x}_{0},
        \hat{\phi}_\text{ATS} = \left(\frac{2E_{C_{J1}+C_{J3}}}{E_{L_\text{ATS}}}\right)^{1/4} \hat{x}_\text{ATS},\; \\
        \hn_{0} &= \left(\frac{E_{L_{0}}}{32E_{C_0}}\right)^{1/4} \hat{y}_{0},\;
        \hn_\text{ATS} = \left(\frac{E_{L_\text{ATS}}}{32E_{C_{J1}+C_{J3}}}\right)^{1/4} \hat{y}_\text{ATS},
    \end{split}
\end{align}
where $\hn = \hQ/2e$ is the cooper pair number operator defined by the charge operator divided by twice the Coulomb charge $e$, $E_{C_i}$ the capacitive energy associated to the capacitance $C_i$ and $E_{L_\text{ATS}} = (E_{L_0}+E_L)$.
We further make the assumption that the Josephson energies are identical $E_{J1} = E_{J3}$, the Hamiltonian becomes
\begin{align}
    \begin{split}\label{eq:Hbm}
        \hH &= \frac{\hbar\omega_0}{4} \left(\hat{x}_0^2 + \hat{y}_0^2 \right) +\frac{\hbar\omega_\text{A}}{4} \left(\hat{x}_\text{ATS}^2 + \hat{y}_\text{ATS}^2 \right) + E_g\hat{x}_0\hat{x}_\text{ATS}\\ & + \hat{x}_0 E^0_\Sigma \phi_\Sigma + \hat{x}_\text{ATS} \left(E^\text{ATS}_\Sigma \phi_\Sigma + E^\text{ATS}_\Delta \phi_\Delta \right) \\
        &- 2E_J\cos(\phi_\Sigma) \cos\left[\varphi_\text{ATS} \hat{x}_\text{ATS}+\phi_\Sigma \frac{c-1}{c+1} \right],
    \end{split}
\end{align}
where we have introduced the notations
\begin{align}
    \begin{split}  
        \varphi_{0} & = \left(\frac{2E_{C_{0}}}{E_{L_{0}}}\right)^{1/4},\\
        \varphi_\text{ATS} & = \left(\frac{2E_{C_{J1}+C_{J3}}}{E_{L_\text{ATS}}}\right)^{1/4},\\
        \omega_0 & = \sqrt{8 E_{C_0} E_{L_0}}, \\
        \omega_\text{A} & = \sqrt{8 E_{C_{J1}+C_{J3}} E_{L_\text{ATS}}}, \\
        E_g & = -\varphi_0 \varphi_\text{ATS} E_{L_0},\\
        E^{0}_\Sigma &= \varphi_0 \frac{2}{c+1} E_{L_0},\\
        E^\text{ATS}_\Sigma &= \varphi_\text{ATS} \left[\frac{-2E_{L_0}}{c+1}+\frac{E_L(c-1)}{c+1}\right],\\
        E^\text{ATS}_\Delta &= -\varphi_\text{ATS} E_L.
    \end{split}
\end{align}

\subsubsection{Normal-mode Hamiltonian }
\label{app:deriv_normal}
The next step is to recast the Hamiltonian in terms of normal modes $a$ and $b$. These modes are obtained by a Bogoliubov transformation, which amounts to diagonalizing the quadratic form on the first row of \cref{eq:Hbm}
\begin{align}
    \begin{split}
        \hat{x}_0 &= u_{0a}\hat{x}_a + u_{0b} \hat{x}_b,\\
        \hat{x}_\text{ATS} &= u_{Aa}\hat{x}_a + u_{Ab} \hat{x}_b,\\
        \hat{y}_0 &= v_{0a} \hat{y}_a + v_{0b} \hat{y}_b,\\
        \hat{y}_\text{ATS} &= v_{Aa} \hat{y}_a + v_{Ab}, \hat{y}_b. \label{eq:Bogoliubov}
    \end{split}
\end{align}
Where the matrices $u$ and $v$ contain the hybridization coefficients that are such that the resulting Hamiltonian has no quadratic cross-terms of the form $\hat{x}_a \hat{x}_b$. The above transformation can be expressed as a function of the circuit parameters. In terms of the new variables on the right-hand side of \cref{eq:Bogoliubov} the Hamiltonian \cref{eq:Hbm} becomes
\begin{align}
    \begin{split}\label{app:eq:H_irrot}
        \hH&= \frac{\hbar\omega_a}{4} \left(\hat{x}_a^2 + \hat{y}_a^2 \right) + \frac{\hbar\omega_b}{4} \left(\hat{x}_b^2 + \hat{y}_b^2 \right)\\
        & + \hat{x}_a \left(E^a_\Sigma \phi_\Sigma + E^a_\Delta \phi_\Delta  \right) + \hat{x}_b \left(E^b_\Sigma \phi_\Sigma + E^b_\Delta \phi_\Delta  \right) \\
        & - 2 E_J \cos(\phi_\Sigma)\cos\left[\varphi_a\hat{x}_a + \varphi_b\hat{x}_b +\phi_\Sigma p_\Sigma \right],
    \end{split}
\end{align}
with,
\begin{align}
    \begin{split}
        E^{a}_\Sigma &= E^0_\Sigma u_{0a} + E^{ATS}_\Sigma u_{Aa},\\
        E^a_\Delta &= E^\text{ATS}_\Delta u_{Aa},\\
        \varphi_a,~ \varphi_b &= \varphi_\text{ATS} u_{Aa},~\varphi_\text{ATS} u_{Ab} \\
        p_\Sigma &= \frac{c-1}{c+1}
    \end{split}
\end{align}
and similarly for the mode $b$. 

For the determination of the normal mode coefficients and frequencies, the transformation that rewrites \cref{eq:Hbm} into \cref{app:eq:H_irrot} can be expressed in terms of the bare mode parameters~\cite{malekakhlagh_lifetime_2020}
\begin{align}
    \begin{split}
        \omega_a &= \sqrt{\omega_0^2 \cos^2\theta +\omega_\text{A}^2\sin^2\theta -2E_g/\hbar\sqrt{\omega_0 \omega_\text{A}} \sin2\theta },\\
        \omega_b &= \sqrt{\omega_\text{A}^2 \cos^2\theta +\omega_0^2\sin^2\theta +2E_g/\hbar\sqrt{\omega_0 \omega_\text{A}} \sin2\theta },\\
        \tan 2\theta &= \frac{4 E_g/\hbar \sqrt{\omega_0 \omega_\text{A}}}{\omega_\text{A}^2 -\omega_0^2}\\
        \begin{bmatrix}
            v_{0a} & v_{0b}\\
            v_{Aa} & v_{Ab}
        \end{bmatrix} & = \begin{bmatrix}
            s_1 s_2 \cos\theta & s_1s_3\sin\theta\\
            -s_1^{-1}s_2 \sin\theta & s_1^{-1}s_3 \cos\theta
        \end{bmatrix},\\
        \begin{bmatrix}
            u_{0a} & u_{0b}\\
            u_{Aa} & u_{Ab}
        \end{bmatrix} & = \begin{bmatrix}
            s_1^{-1} s_2^{-1} \cos\theta & s_1^{-1}s_3^{-1}\sin\theta\\
            -s_1s_2^{-1} \sin\theta & s_1s_3^{-1} \cos\theta
        \end{bmatrix},\\
        s_1,s_2,s_3 &= \left(\frac{\omega_\text{A}}{\omega_0}\right)^{1/4},~ \left(\frac{\omega_a^2}{\omega_0\omega_\text{A}} \right)^{1/4},~ \left(\frac{\omega_b^2}{\omega_0\omega_\text{A}} \right)^{1/4}.
    \end{split}
\end{align}

\subsubsection{Displacement transformation}
\label{app:deriv_disp}
We now consider the following form for the time-dependent external fluxes in the time-dependent Hamiltonian \cref{app:eq:H_irrot}
\begin{align}
    \begin{split}\label{eq:varphiDelSig}
        \phi_\Delta(t) &= \phi_\Delta^0 + \left( -i\eta_p/2 e^{i\omega_pt} +\text{c.c}\right),\\
        \phi_\Sigma(t) &= \phi_\Sigma^0 + \left(-i\epsilon_p/2 e^{i\omega_pt } + \text{c.c}\right).
    \end{split}
\end{align}
To bring the Hamiltonian to the form in the main text \cref{eq:H-circuit_1}, we want to cancel the drive term in the cosine potential of \cref{app:eq:H_irrot} with a displacement transformation $\mathcal{D}_c = \exp\left[i\left( y_c^{disp} \hat{x}_c - x_c^{\textit{disp}} \hat{y}_c \right)\right]$ with $c = a, b$, chosen such that it satisfies
\begin{align}
    \begin{split}\label{app:eq:constraints_displacement}
        x_a^{disp} \varphi_a + x_b^{disp} \varphi_b &= - \phi_\Sigma p_\Sigma - \pi/2,\\
        \dot{x}_a^{disp} &= \frac{\omega_a}{2} y_a^{disp},\\
        \dot{x}_b^{disp} &= \frac{\omega_b}{2} y_b^{disp},\\
        \varphi_b \Big(E_\Sigma^a\phi_\Sigma + E_\Delta^a \phi_\Delta \hspace{20pt}&\hspace{10pt} \varphi_a\Big(E_\Sigma^b\phi_\Sigma + E_\Delta^b \phi_\Delta \\
        + \frac{\hbar\omega_a}{2} x_a^{disp} + \dot{y}_a^{disp}  \Big)&= \hspace{20pt} + \frac{\hbar\omega_b}{2} x_b^{disp} + \dot{y}_b^{disp}  \Big).
    \end{split}
\end{align}
The first condition imposes that the cosine in \cref{app:eq:H_irrot} is transformed to $\sin(\hphi_a p_a + \hphi_b p_b)$. The second and third equations impose that there is no linear term in the charge. The last equation enforces that the inductive drive terms are of the form $E_L^{\textit{eff}} (\hat{x}_a \varphi_a + \hat{x}_b \varphi_b)$. One can easily solve these equations with the following Ansatz
\begin{align}
\begin{split}
    x_{a/b}^{disp} &= x_{a/b}^0 + \frac{1}{2i}\left(x_{a/b}^p e^{i\omega_p t} - x_{a/b}^{p*} e^{-i\omega_pt}\right),\\
    y_{a/b}^{disp} &= y_{a/b}^p e^{i\omega_p t} + y_{a/b}^{p*} e^{-i\omega_p t},
\end{split}
\end{align}
where $x_{a}^{p*}$ is the complex conjugate of $x_a^p$ and likewise for $x_b^p,y_a^p$ and $y_b^p$. The solutions read
\begin{widetext}
    \begin{align}
        \begin{split}
            x_a^0 &= \frac{ - \omega_b p_{\Sigma} \varphi_{a} \phi^{0}_{\Sigma} - \pi/2 \omega_b \varphi_{a} - 2 E_{\Delta}^a/\hbar \varphi_{b}^{2} \phi^{0}_{\Delta} + 2 E_{\Delta}^b/\hbar \varphi_{a} \varphi_{b} \phi^{0}_{\Delta} - 2 E_{\Sigma}^a/\hbar \varphi_{b}^{2} \phi^{0}_{\Sigma} + 2 E_{\Sigma}^b/\hbar \varphi_{a} \varphi_{b} \phi^{0}_{\Sigma}}{\omega_a \varphi_{b}^{2} + \omega_b \varphi_{a}^{2}},\\
            x_b^0 &= \frac{ - \omega_a p_{\Sigma} \varphi_{b} \phi^{0}_{\Sigma} - \pi/2 \omega_a \varphi_{b} + 2 E_{\Delta}^a/\hbar \varphi_{a} \varphi_{b} \phi^{0}_{\Delta} - 2 E_{\Delta}^b/\hbar \varphi_{a}^{2} \phi^{0}_{\Delta} + 2 E_{\Sigma}^a/\hbar \varphi_{a} \varphi_{b} \phi^{0}_{\Sigma} - 2 E_{\Sigma}^b/\hbar \varphi_{a}^{2} \phi^{0}_{\Sigma}}{\omega_a \varphi_{b}^{2} + \omega_b \varphi_{a}^{2}},\\
            x_a^p =&\frac{\epsilon_p\left[ 2E^a_\Sigma/\hbar \varphi_b^2-2E^b_\Sigma/\hbar\varphi_a\varphi_b + (\omega_b -4\omega_p^2/\omega_b)p_\Sigma \varphi_a \right] +
            \eta_p\left[2E^a_\Delta/\hbar\varphi_b^2- 2E^b_\Delta/\hbar\varphi_a\varphi_b \right] }{(\omega_a- 4\omega_p^2/\omega_a)\varphi_b^2 + (\omega_b -4\omega_p^2/\omega_b)\varphi_a^2},\\
            x_b^p =&\frac{\epsilon_p\left[ 2E^b_\Sigma/\hbar \varphi_a^2-2E^a_\Sigma/\hbar\varphi_a\varphi_b + (\omega_a -4\omega_p^2/\omega_a)p_\Sigma \varphi_b \right] +
            \eta_p\left[2E^b_\Delta/\hbar\varphi_a^2- 2E^a_\Delta/\hbar\varphi_a\varphi_b \right] }{(\omega_a- 4\omega_p^2/\omega_a)\varphi_b^2 + (\omega_b -4\omega_p^2/\omega_b)\varphi_a^2},\\
            y_{a/b}^p &= \frac{\omega_p}{\omega_{a/b}} x_{a/b}^p.
        \end{split}
    \end{align}
\end{widetext}

After this displacement, the Hamiltonian has the following expression
\begin{align}\label{app:eq:H_syst}
    \begin{split}
        \hH_s =& \frac{\hbar\omega_a}{4} \left( \hat{x}_a^2 + \hat{y}_a^2 \right) + \frac{\hbar\omega_b}{4}\left(\hat{x}_b^2 + \hat{y}_b^2 \right)\\
        &-E_L^{\textit{eff}}(\epsilon_p,\eta_p)  \left[\frac{e^{i\omega_p t}}{2i} + \text{c.c.}\right] \left(\varphi_a \hat{x}_a + \varphi_b \hat{x}_b \right)\\
        &+E_L^0 \left( \varphi_a \hat{x}_a  +  \varphi_b \hat{x}_b  \right)\\
        & - 2 E_J \cos(\phi_\Sigma)\sin\Big(\varphi_a \hat{x}_a + \varphi_b \hat{x}_b  \Big),
    \end{split}
\end{align}
with
\begin{align}
    \begin{split} \label{eq:EL0ELeff}
        E_L^0 &= \frac{\hbar\omega_a x^{0}_{a} + 2E_{\Delta}^a \phi^{0}_{\Delta} + 2E_{\Sigma}^a \phi^{0}_{\Sigma}}{2\varphi_{a}}, \\
        E_L^{\textit{eff}}(\epsilon_p,\eta_p) &= \frac{-\hbar\omega_a x^{p}_{a} -2 E_{\Delta}^a \eta_{p} -2 E_{\Sigma}^a \epsilon_{p} + 4y^{p}_{a} \hbar\omega_{p}}{2\varphi_{a}}.
    \end{split}
\end{align}
Finally, we are free to tune the DC phases $\phi_\Sigma^0$ and $\phi_\Delta^0$~\cite{lescanne_exponential_2020}. We take $\phi_\Sigma^0 = -\pi/2$ and $\phi_\Delta^0$ is chosen such that $E_L^0=0$ in \cref{eq:EL0ELeff}. When additionally choosing the ratio of the pump amplitudes in \cref{eq:varphiDelSig} $\eta_p/\epsilon_p$ such that $E_L^{\textit{eff}}=0$ in the second \cref{eq:EL0ELeff}, we obtain the Hamiltonian \cref{eq:H-circuit_1}. 

In \cref{eq:H-circuit_1} an additional charge drive has been added on the node $4$ resulting in a term $\epsilon_d \cos(\omega_dt) (\hat{y}_b + u\hat{y}_a)$, with $u=v_{4a}/v_{4b}$. In \cref{sec:mitigation}, we relax the constraint on the ratio $\eta_p/\epsilon_p$. We argue that the Hamiltonian in \cref{sec:mitigation} is still captured by the effective model derived in \cref{sec:eff-mod}, upon the redefinition of $\mathrm{g}_{1,1}$. Finally, note that the obtained Hamiltonian \cref{eq:H-circuit_1} is quite generic for cat state stabilization using an ATS non-linear element~\cite{lescanne_exponential_2020, berdou_one_2023}.

\subsubsection{Josephson junction asymmetries}
\label{app:deriv_asymm}
In realistic circuits, the Josephson junctions in \cref{app:eq:H_syst} are not symmetric. Let us denote their corresponding Josephson energies by $E_{J1}$ and $E_{J3}$. This results in an additional term in \cref{app:eq:H_syst} of the form $\Delta E_J \sin(\phi_\Sigma)\cos(\hat{x}_a \varphi_a+ \hat{x}_b \varphi_b)$, with $\Delta E_J = \left(E_{J1}-E_{J3}\right)/2$ and the redefinition $E_J = \left(E_{J1} + E_{J3}\right)/2$. Experimentally, this asymmetry is such that $\Delta E_J = 0.01 E_J = O(\lambda^2E_J)$. When expanding the non-linear potential $\Delta E_J \cos(\epsilon)\cos(\hat{x}_a \varphi_a+ \hat{x}_b \varphi_b)$ the leading  term is of order $O(\lambda^6 E_J)$ (see \cref{app:trunc}). The associated off-resonant dressing will therefore be negligible contributions to the rates computed in \cref{fig:Collapse_monom}.

\subsubsection{Tolerance on the DC flux imprecision}
\label{app:deriv_flux_tol}
Under \cref{eq:EL0ELeff}, we have set the DC fluxes $\phi_\Sigma^0=-\pi/2$ and $\phi_\Delta^0$ such that $E_L^0=0$. We will analyze the impact of a miscalibration of the DC fluxes. We introduce $\delta \varphi$ so that the DC parts of the flux drives are now $ \delta\varphi + \phi_\Delta^0$ and $\delta\varphi + \phi_\Sigma^0$. Starting from \cref{app:eq:H_syst}, we get the following Hamiltonian to first order in $\delta\varphi$,
\begin{align}\label{app:eq:H_error}
    \begin{split}
        \hH_s =& \frac{\hbar\omega_a}{4}\left(\hat{x}_a^2 + \hat{y}_a^2 \right) + \frac{\hbar\omega_b}{4} \left(\hat{x}_b^2 + \hat{y}^2 \right)\\
        &+ \left[E_L^{\textit{eff}}(\epsilon_p,\eta_p) e^{i\omega_p t} +\text{c.c.} \right]\left( \varphi_a \hat{x}_a  + \varphi_b \hat{x}_b  \right) \\
        &+(E_\Delta^a+E_\Sigma^a)\delta\varphi \hat{x}_a  +(E_\Delta^b+E_\Sigma^b)\delta\varphi \hat{x}_b \\
        & - 2 E_J \cos(\phi_\Sigma)\sin\Big(\varphi_a \hat{x}_a + \varphi_b \hat{x}_b  \Big)\\
        & + 2E_J \sin(\phi_\Sigma) \delta\varphi \sin\Big(\varphi_a \hat{x}_a + \varphi_b \hat{x}_b  \Big).
    \end{split}  
\end{align}
Additionally displacing $\hat{x}_a$ by ${-2\delta\varphi (E_\Delta^a+E_\Sigma^a)/\omega_a \doteq \delta\varphi p_\delta^a}$ and similarly for $\hat{x}_b$. To lowest order in $\delta \varphi$ we get two contributions 
\begin{align}
    \begin{split}
        2E_J\sin[\epsilon(t)] (p_\delta^a + p_\delta^b)\delta\varphi \cos\Big(\varphi_a \hat{x}_a + \varphi_b \hat{x}_b  \Big)\\
        +2E_J\cos[\epsilon(t)]\delta\varphi\sin\Big(\varphi_a \hat{x}_a + \varphi_b \hat{x}_b  \Big).
    \end{split}
\end{align}
To lowest order in $\lambda$ both of these terms are of the order of $O(E_J \lambda^3 \delta\varphi)$ (see \cref{app:trunc}), where we assumed $p_\delta^a, p_\delta^b= O(1)$. We conclude that we can tolerate an imprecision of the DC fluxes up to $\lambda^3 \Phi_0$, with $\Phi_0$ the flux quantum. 

\begin{widetext}
\subsection{Comparing Flux and Charge bath couplings}
In this appendix, we argue that flux noise, although present in our system, is in practice negligible by comparison to charge noise. This appendix supports the assumptions presented in \cref{sec:eff-mod}, in particular the modeling of the undriven system-bath coupling.
\label{app:bathcoupl}

To model the coupling of the system to a noise source in the external flux, we take the derivative of the lab-frame Hamiltonian under the irrotational constraint in \cref{app:eq:H_irrot} with respect to the external fluxes
\begin{align}
    \begin{split}
        \frac{\partial \hH}{\partial \phi_\Delta} =& E_\Delta^a \hat{x}_a + E_\Delta^b \hat{x}_b\\ 
        \frac{\partial \hH}{\partial \phi_\Sigma} =& E_\Sigma^a \hat{x}_a + E_\Sigma^b \hat{x}_b + p_\Sigma 2 E_J \cos(\phi_\Sigma)\sin\left(\varphi_a \hat{x}_a + \varphi_b \hat{x}_b+ p_\Sigma \phi_\Sigma \right)\\
        &\hspace{60pt}+ 2 E_J \sin(\phi_\Sigma)\cos\left(\varphi_a \hat{x}_a + \varphi_b \hat{x}_b + p_\Sigma \phi_\Sigma \right).
    \end{split}
\end{align}
After applying the displacement transformation used to obtain the circuit Hamiltonian in \cref{app:eq:H_syst} we get the following coupling to the flux line, treated as a quantum noise source
\begin{align}
    \begin{split}
        \hH_{\textit{sf1}} &= \left[ E_\Delta^a (\hat{x}_a + x_a^\textit{disp}) + E_\Delta^b (\hat{x}_b + x_b^\textit{disp}) \right] \otimes \hat{B}_{\textit{f1}}(t),\\
        \hH_{\textit{sf2}} &= \left[ E_\Sigma^a (\hat{x}_a + x_a^\textit{disp}) + E_\Sigma^b (\hat{x}_b + x_b^\textit{disp}) - p_\Sigma 2 E_J \sin[\epsilon(t)]\cos\left(\varphi_a \hat{x}_a + \varphi_b \hat{x}_b \right)- 2 E_J \cos[\epsilon(t)]\sin\left(\varphi_a \hat{x}_a + \varphi_b \hat{x}_b \right)\right] \otimes  \hat{B}_{\textit{f2}}(t),
    \end{split}
\end{align}
where we have denoted $\phi_\Sigma = -\frac{\pi}{2} + \epsilon $. To lowest order in $\lambda$, we can take the flux coupling to be of the form $\left(E_\Delta^a \hat{x}_a + E_\Delta^b \hat{x}_b\right)\otimes \hat{B}_{f1} + \left[(E_\Sigma^a-2E_Jp_a) \hat{x}_a + (E_\Sigma^b-2E_Jp_b) \hat{x}_b\right]\otimes \hat{B}_{f2} $. The next terms are of order $\lambda^3$. The obtained flux coupling is time-independent and does not directly probe the low-frequency flux noise. 

Contrary to the charge coupling that has to remain large enough to satisfy the adiabatic elimination condition [see \cref{eq:adiabatic_condition}], the flux-noise can be engineered to be small by adjusting circuit parameters. Therefore, we assume that the pump-activated mechanisms coming from the flux coupling are negligible. 
\end{widetext}

\section{Truncation scheme for the effective model}\label{app:trunc}
The expansions presented in this work are all performed with respect to a parameter $\lambda$ that is of the same order as $\epsilon_p, \varphi_a, \varphi_b$. In this section, the interplay between the truncation of the system Hamiltonian  \cref{eq:H-circuit_1} and the order of the SWPT that captures all the contributions below the truncation order is discussed. 
\Cref{app:sec:truncRWA} is used in \cref{sec:eff-mod} where it allows for deriving the effective Hamiltonian to order $E_J \lambda^6$. \Cref{app:sec:expansions} is used to perform the normal-ordered expansions of \cref{eq:H-circuit_1} and describes the Husimi-Q phase-space representation on which the symbolic algorithm that performs the SWPT relies.

\subsection{Truncation and SWPT}\label{app:sec:truncRWA}
We start with the formulas for the SWPT derived in~\cite{venkatraman_static_2022}
\begin{align}
    \hat{K}_{[k]}^{(n)} &= \begin{cases}
        \hH_s & n=k=0\\
        \dot{\hat{S}}^{(n+1)} + \frac{1}{i\hbar}\left[ \hat{S}^{(n)},H_s \right] & k=1 \\
        \sum\limits_{m=0}^{n-1} \dfrac{1}{ki \hbar} \left[\hat{S}^{(n-m)}, \hat{K}_{[k-1]}^{(m)} \right] & 1<k\leq n+1 
    \end{cases} \\
    \hat{S}^{(n+1)} &=  \begin{cases}
        - \int \operatorname{osc}(\hH_s) dt & n=0\\
        - \int dt \operatorname{osc} \Big( \dfrac{1}{i\hbar}\left[ \hat{S}^{(n)},\hH_s \right] \\+ \sum\limits_{k>1}^{n+1}\sum\limits_{m=0}^{n-1} \dfrac{1}{ki \hbar} \left[\hat{S}^{(n-m)}, \hat{K}_{[k-1]}^{(m)} \right]  \Big) & n>0, 
    \end{cases}  
\end{align}
where we define the oscillatory part of a time-dependent operator as,
\begin{align} \label{eq:osc}
    \mathrm{osc} \left( \hat{O}(t) \right) = \hat{O}(t) - \lim_{T\to\infty} \int_0^T dt \hat{O}(t)/T.
\end{align}
The aim of this section is to use these formulas to analyze under which circumstances the procedure converges. Convergence should be understood in the sense that given a truncation order $E_J\lambda^k$ of the system Hamiltonian $\hH_s$, there exists an iteration number $n$ such that any contribution coming from the next iterations of the SWPT will be of order larger than $k$.
\FloatBarrier
\begin{figure}[!htbp]
    \centering
    \includegraphics[width=0.8\linewidth]{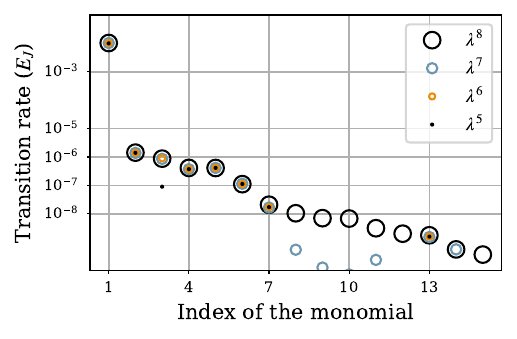}
    \caption{Ranking of the monomials used in \cref{fig:Collapse_monom}: For each monomial, the corresponding transition rate between Fock states is represented in units of $E_J$. We show various truncation orders and associated SWPT to highlight the convergence of the calculated rates. Note that at truncation and SWPT $\lambda^n$, we obtain a precise result up to $E_J\cdot 10^{-n}$ in the rates, since $\lambda\simeq10^{-1}$ numerically.}
    \label{fig:enter-label}
\end{figure}

We define the time-dependent Hamiltonian at iteration $n$ for $n>0$.
\begin{align}\label{eq:trunc_Hn}
    \hH^{(n)}(t) &= \left[\frac{ \hat{S}^{(n)}}{i\hbar},\hH_s \right] +
        \sum\limits_{k>1}^{n+1}\sum\limits_{m=0}^{n-1} \dfrac{1}{k} \left[\frac{ \hat{S}^{(n-m)}}{i\hbar}, \hat{K}_{[k-1]}^{(m)} \right].
\end{align}
The effective Hamiltonian at iteration $n$ is $\hat{K}^{(n)} = \sum_k K_{[k]}^{(n)} = \overline{H^{(n)}(t)} = \hH^{(n)}(t) - \mathrm{osc}(\hH^{(n)}) $. We have the following relations for the order of each quantity
\begin{align*}
    \left|\hH_s\right| &= O(E_J \lambda^2),\\
    \left|\overline{H_s}\right| &= O(E_J \lambda^4),\\
    \left|\hat{K}^{(n)}\right| &= O\left(\left|\hH^n\right|\right),\\
    \left|\hat{K}_{[k]}^{(n)}\right| &= O\left(\left|\hH^n\right|\right). 
\end{align*}
Using the previous equations, we can derive the order of $\left|\hat{S}^{(n+1)}\right| = O\left(\frac{\left|\hH^{(n)}\right|}{\omega - \omega^*}\right)$, with $\omega-\omega^*$ the frequency of the process, this notation highlights that the SWPT is ill-defined for arbitrary slowly rotating contributions.
Finally, using \cref{eq:trunc_Hn} and proceeding by induction one obtains the order of the time-dependent Hamiltonian at the $n^{\text{th}}$ iteration
\begin{align}\label{app:eq:orderHn}
    \left|\hH^{(n)}\right| &= O\left( \frac{E_J^{n+1} \lambda^{2(n+1)}}{\prod_{i=1}^n \hbar(\omega-\omega_i^*)} \right).
\end{align}

From \cref{app:eq:orderHn} it is clear that one can obtain a given precision after any finite number of SWPT iterations provided that $ \frac{E_J^{n+1} \lambda^{2(n+1)} }{\prod_{i=1}^n \hbar(\omega-\omega_i^*)} $ goes to $0$ when $n$ increases. In general, the previous quantity does not decrease with $n$ since near-resonant contribution at order $n$ can be of arbitrary magnitude, as highlighted by the notation $\frac{1}{\omega-\omega_i*}$. 

We make the assumption $\prod_{i=1}^n \hbar (\omega-\omega_i^*)>\lambda^n E_J^n$
which ensures that $|\hat{H}^{(n)}| = O(E_J \lambda^{n+2})$. 
Therefore, when truncating the starting-point $\hat{H}_s$ to order $\lambda^k$, only the $k-2$ first iterations of the SWPT procedure will possibly give contributions above $E_J \lambda^k$. With this assumption, we choose the number of SWPT iterations along with the order at which we truncate the Josephson potential in \cref{eq:Hs_g_ology}.

The above assumption is motivated by high-frequency expansions \cite{venkatraman_static_2022} and $\hbar\omega_a, \hbar\omega_b~>~E_J\lambda$ in typical circuit implementations. However, as mentioned above, near-resonant contributions will violate our assumption. In the case of a near-resonant contribution, the effective model will be obtained via adiabatic elimination methods, resulting in a denominator $1/\kappa_b$. For typical dissipative-cat stabilization, $\kappa_b$ is large by design and of the order of $E_J\lambda^2 $. As a result, a near-resonant contribution can lead to an underestimation of the error in the truncation and SWPT procedure described above. Due to the magnitude of $\kappa_b$, we expect bounded corrections from near-resonant contributions that can be captured with higher order truncations and SWPT procedures.

We choose the truncation of the system Hamiltonian \cref{eq:Hs_g_ology} such that $\kappa_1 \simeq E_J\lambda^k$, where $k$ is the order at which the system Hamiltonian is expanded. In \cref{sec:eff-mod} we set $k=6$.

\subsubsection*{Explicit formula for the system-bath coupling}
In this section, we give the formula to calculate the system-bath interaction at a given order in the SWPT. We expand the system-bath coupling in orders of the perturbation parameter and write $e^{\hat{S}/i\hbar}\hH_{\textit{sB}}e^{-\hat{S}/i\hbar} = \sum_n \hH_\textit{sB}^{(n)} $. With a similar reasoning to the one presented in~\cite{venkatraman_static_2022} we introduce the quantities $\hH_{\textit{sB}}^{(n)} = \sum_k \hH^{(n)}_{[k]}$ and obtain 
\begin{align}
    \hH^{(n)}_{[k]} &=  \begin{cases}
        \hH_\textit{sB} & n=0=k,\\
        \left[\frac{\hat{S}^{(n)}}{i\hbar}, \hH_\textit{sB}\right] & k=1,\\
        \sum\limits_{m=0}^{n-1} \dfrac{1}{k} \left[\frac{\hat{S}^{(n-m)}}{i\hbar}, \hH^{(m)}_{[k-1]} \right] & 1<k\leq n+1.
    \end{cases} 
\end{align}

Using the SWPT we have obtained an effective, time-independent system Hamiltonian and a time-dependent system-to-bath coupling Hamiltonian. We can now construct the master equation by tracing out the bath.

% In this section, we give the formula to calculate the system-bath interaction at a given order in the SWPT. We expand the system-bath coupling in orders of the perturbation parameter and write $\hH_{\textit{sB}} = \sum_n \hH_\textit{sB}^{(n)} $. With a similar reasoning to the one presented in~\cite{venkatraman_static_2022} we obtain 
% \begin{align}
%     \hH_\textit{sB}^{(n)} &=  \begin{cases}
%         \hH_\textit{sB} & n=0,\\
%         \left[\frac{\hat{S}^{(n)}}{i\hbar}, \hH_\textit{sB}\right] + 
%         \sum\limits_{m=0}^{n-1} \dfrac{1}{n} \left[\frac{\hat{S}^{(n-m)}}{i\hbar}, \hH_\textit{sB}^{(m)} \right] & n>0.
%     \end{cases} 
% \end{align}
% This formula allows for a fast implementation of the symbolic algorithm.

\subsection{Normal ordered expansion of the trigonometric functions}\label{app:sec:expansions}
All the calculations presented in this work are performed using normal-ordered operators. A very convenient way to calculate the product of normal-ordered operators is to use the Husimi-Q representation~\cite{carmichael_statistical_2008}. In particular, expressions containing operators are related to their Husimi-Q function counterpart by the following relation,
\begin{align*}
    F(\ha , \ha^\dagger) \overset{\text{Normal ordering}}{\to } G(\ha,\ha^\dagger)  \overset{\text{Husimi-Q function}}{\to }G(\alpha,\alpha^*)
\end{align*}
where $F$ is an arbitrary function and $G$ the function obtained when writing $F$ in normal order and $\alpha$ is the phase-space variable that is a complex number.

From here we can easily calculate the Husimi-Q function of the sine non-linearity of \cref{eq:H-circuit_1}
\begin{align}
\begin{split}
    \exp\left[ i\varphi_a ( \ha +\ha^\dagger ) \right] &= e^{-\varphi_a^2/2} e^{i\varphi_a \ha^\dagger }e^{i\varphi_a \ha }, \\
    &\to e^{-\varphi_a^2/2} e^{i\varphi_a \alpha^* }e^{i\varphi_a \alpha },\\  
    \sin\left[\varphi_a (\ha+\ha^\dagger )\right]& \to e^{-\varphi_a^2/2} \sin\left[ \varphi_a (\alpha + \alpha^*)\right].
\end{split}
\end{align}
The remaining $\sin\left[ \varphi_a (\alpha + \alpha^*)\right]$ is Taylor expanded in the small parameter $\varphi_a$. The main advantage of performing the symbolic calculations in the Husimi phase-space is that one can recast the calculation of the nested commutators appearing in the SWPT by derivatives of the phase-space functions \cite{carmichael_statistical_2008}. Moreover, these phase-space functions will be only polynomials. Using the built-in features of the package \textit{sympy}~\cite{meurer_sympy_2017} partial derivatives of polynomials are efficiently computed.

Using the above formalism \cref{eq:Hs_g_ology} can be rewritten
\begin{align}
\begin{split} \label{app:eq:Hs_g_ology_phase}
    H_s/\hbar =& \delta |\alpha|^2 + \Delta |\beta|^2 -i \epsilon_d \cos(\omega_d t )  (\beta e^{-i\omega_d t}-\beta^* e^{i\omega_d t} )\\
    &  +\sum_{n,k\ \text{odd}} {\mathrm{g}}_{n,k}\ e^{ik\omega_p t} \left[x_a(t) + rx_b(t) \right]^{n} +\text{h.c.},
\end{split}
\end{align}
where $x_\eta(t)$ is the phase-space function associated to $\hat{x}_{\eta}(t)$ for $\eta=a, b$. Note that the same renormalization would have been obtained when expanding the Hamiltonian in normal-ordered operators. Moreover, this renormalization increases the range of validity of the model for $\varphi_{a,b}$, similar to the Jacobi-Anger expansion.
\FloatBarrier
\begin{figure}[t!]
    \centering
    \includegraphics[width=\linewidth]{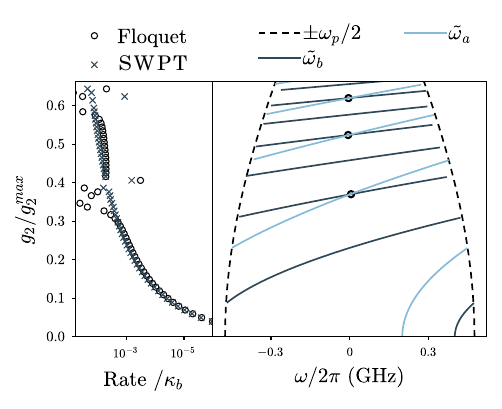}
    \caption{Validity at large pump amplitudes: on the left-hand side, decay rate from the sector $N_d:1\to0$ (see \cref{fig:G_a-0_eps}). For large $g_2/g_2^\textit{max}$, the system is ac Stark shifted on resonance. On the right-hand side, the effective model is used to predict the ac Stark shift of the modes. The frequencies $\tilde{\omega}_a, \tilde{\omega}_b$ are folded in the first Brillouin zone. This amounts to taking the remainder upon division by $\omega_p$, which depends on the pump power and is the solution of \cref{eq:pump_match}. When the two lines cross, the modes undergo a pump-assisted resonance if the selection rules of the system allow it.
    }
    \label{fig:Res}
\end{figure}

\section{Agreement between SWPT and Floquet numerical simulations}
\label{app:agreement}

In this section, we study the agreement between SWPT and Floquet theory (see \cref{fig:G_a-0_eps} in \cref{sec:Comparison}) more in depth.

In \cref{fig:Res}, the single-photon loss rate of the mode $a$ (transition from sector $N_d:1\to0$ \cref{fig:G_a-0_eps}) is shown for larger pump $g_2/g_2^\textit{max}$. For $g_2/g_2^\textit{max}$ close to $0.4$, we observe a divergence in the Floquet rates. On the right of \cref{fig:Res}, we analyze the origin of this peak using the effective Hamiltonian \cref{eq:Heff_6}. We find that it is the result of the pump ac Stark shifting the frequencies of the two normal modes $a$ and $b$ into resonance.  
To understand this, the blue solid lines in \cref{fig:Res} show the ac Stark shifted mode frequencies $\tilde{\omega}_a, \tilde{\omega}_b$ of \cref{eq:res-cond}, in the first Brillouin zone. The dashed black lines represent the boundaries of the first Brillouin zone, $\pm \omega_p/2$, as a function of $g_2/g_2^\textit{max}$, where the dependence arises from the resonance matching condition \cref{eq:pump_match}. 
\begin{figure*}
    \centering
    \includegraphics[width=\textwidth]{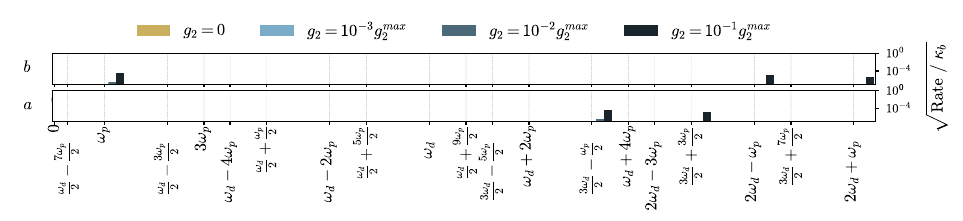}
    \caption{Difference between \cref{fig:Collapse_monom} where $\epsilon_d=5g_2$ and the same plot with $\epsilon_d=0$. The minimum of the $y$-axis is $10^{-7}$. We see that $\epsilon_d$ is accountable for contributions which result in rates below $10^{-8}\kappa_b$.}
    \label{fig:Collapse_monom_ed0}
\end{figure*}
When the mode frequencies match up to an integer multiple of the pump frequency, there is a pump-mediated resonance, with the corresponding crossing highlighted by a circle in \cref{fig:Res}. The pump powers at which these resonances appear agree with the powers at which divergences occur. In the vicinity of divergences, there is therefore strong hybridization that causes state mistracking, so in \cref{fig:G_a-0_eps} we cut the range of pump power right after the first divergence. By counting the number of folding into the first Brillouin zone we get the resonance condition, $\tilde{\omega}_a = \tilde{\omega}_b -4\omega_p$. Since there is no rotating term of the form $\ha \hb^\dagger$ in \cref{eq:H-circuit_1},  we expect this contribution to be at least of second order in the SWPT. Several second-order processes satisfy this resonance condition, among which the lowest order in $\lambda$ is $g_{1,5}\ha^\dagger$, which is already of order $\lambda^6$. The effect of this term in the second order in SWPT and the subsequent resonance is therefore not captured by the effective model. This discrepancy would be diminished if strong buffer dissipation as required by the adiabaticity condition in \cref{eq:adiabatic_condition} were included in the Floquet simulations.
\FloatBarrier
\begin{figure*}%[t!]
    \centering
    \includegraphics[width=\textwidth]{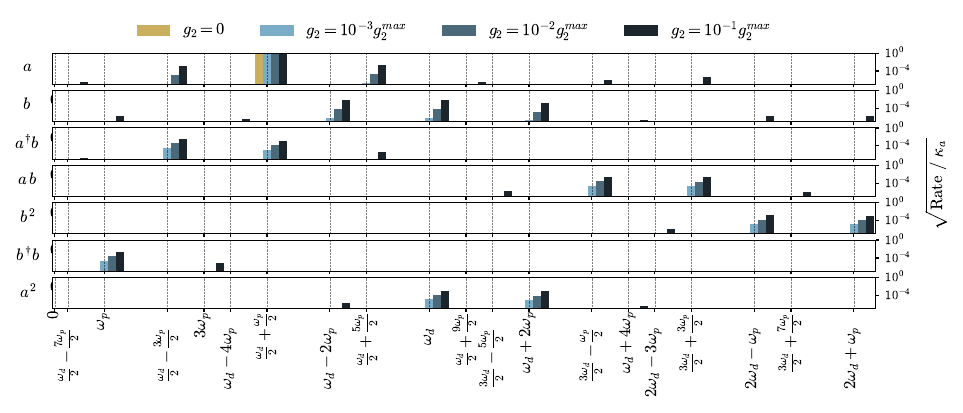}
    \caption{Analysis of the off-resonant dressing of the system-bath coupling to the mode $a$: Absolute value of the prefactor of the monomial on the $y$-axis in the collapse operator $C(\omega_i)$ whose frequency $\omega_i$ is given on the $x$-axis. The form of the coupling is related to the dissipators in the effective master equation \cref{eq:L_eff}. In this figure, we set $\epsilon_d = 5g_2$.}
    \label{fig:Collapse_monom_a}
\end{figure*}

\begin{figure}%[t!]
    \centering
    \includegraphics[width=\linewidth]{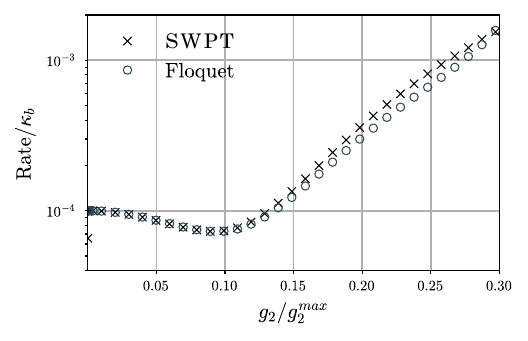}
    \caption{Single-photon loss rate of the mode $a$ as defined in \cref{fig:Fl-tran-matrix} between $N_d:0\to 1$. The hybridization of the modes is taken into account with $u=0.01$ (see \cref{eq:H-circuit_1}). We observe two distinct monotonicities, decreasing when the undressed decay rate of the mode $a$ dominates and increasing when the dressed decays of the mode $b$ dominate. Parameters as in \cref{fig:G_a-0_eps}.}
    \label{fig:Rate_na_nb}
\end{figure}
\section{Additional decay channel analysis}\label{app:extra_collapse}

In this Appendix, we discuss the effect of the linear charge drive $\epsilon_d$ on the decay rates derived in \cref{fig:Collapse_monom}. We also discuss the off-resonant dressing of the mode $a$ and find similar spectral features of the dressings.

\subsection{Effect of \texorpdfstring{$\epsilon_d$}{epsilon d}}\label{app:extra_collapse:ed}
As discussed in \cref{sec:eff-mod:results}, the off-resonant dressing of the system-bath coupling stemming from the drive $\epsilon_d$ is expected to be small since $\epsilon_d = O(E_J \lambda^4)$. In \cref{fig:Collapse_monom_ed0}, the amplitude of the prefactors of the various monomials in the collapse operators for $\epsilon_d=0$ is shown. The difference between \cref{fig:Collapse_monom} and \cref{fig:Collapse_monom_ed0} is smaller than $10^{-4}$, and we conclude therefore that the effect on the effective decay rates is negligible.

\subsection{Off-resonant dressing of the mode \texorpdfstring{$a$}{a}}
\label{sec:aToBath}
In \cref{fig:Collapse_monom_a}, we analyze the collapse operators resulting from the off-resonant dressing of the system-bath coupling of the mode $a$. We report similar spectral features for the parity-breaking effective dissipators, $\ha$, $\ha\hb$, $\ha^\dagger \hb$ to the one identified in \cref{fig:Collapse_monom}. Moreover, the scaling with respect to $g_2/g_2^\textit{max}$ is of the same order as in \cref{fig:Collapse_monom}. The importance of the off-resonant dressings of mode $a$ compared to the one of mode $b$ depends on the factor $u$ which is computed with microwave simulations. The fact that the spectral features are similar to the ones of \cref{fig:Collapse_monom} ensures that filtering the spurious decay processes as in \cref{sec:mitigation} will also mitigate the spurious decays stemming from the system-bath coupling of mode $a$.

In \cref{fig:Rate_na_nb} we perform the Floquet analysis presented in \cref{sec:strong-drive} for a non-zero coupling of the mode $a$ to the bath. We focus on the single-photon loss rate of the mode $a$. We set the ratio of the hybridization coefficients to $u=0.01$ [see \cref{eq:H-circuit_1}]. At small $g_2/g_2^\textit{max}$, we have the expected Purcell rate, without filtering of the transmission line we have $\kappa_a = u^2 \kappa_b$. When increasing $g_2/g_2^\textit{max}$, the rate decreases and for $g_2/g_2^\textit{max}>0.1$ it starts increasing. This second regime is the one where the dressed decays of the mode $b$ are dominating.
A typical experimental value is $u=0.06$. \Cref{fig:Rate_na_nb} was obtained assuming no filters on the transmission line. The value $u=0.01$ was chosen so to obtain a change in monotonicity of the rate before the first resonance \cref{app:agreement}. This value is consistent with typical experimental value of $\kappa_a/\kappa_b = 10^{-4} - 10^{-5}$.
 
In this appendix, we analyzed the decay channels stemming from the mode $a$ and established that the charge drive $\epsilon_d$ has a negligible contribution to the decay rates presented in \cref{fig:Collapse_monom}.  
%\FloatBarrier

\section{Derivation of the effective master equation}\label{app:sec:eff-Lind}
In this section we detail the derivation of the effective Lindblad master equation, starting from \cref{eq:HsbI_collapse} to obtain \cref{eq:L_eff}. We will introduce the secular, Born and Markov approximations that are required for the derivation of a Lindblad master equation~\cite{breuer_dissipative_1997, petrescu_lifetime_2020}.

In \cref{eq:HsbI_collapse} we go into the interaction picture with respect to the effective Hamiltonian $\hat{K}$,
\begin{align}
\begin{split}
    \hH_\textit{sB}^\textit{I} &= e^{i\hat{K}t/\hbar}e^{\hat{S}(t)/i\hbar}\hH_\textit{sB}e^{-\hat{S}(t)/i\hbar}e^{-i\hat{K}t/\hbar}, \\
    &= \hbar \sum_{j,k} \hat{C}^{(k)}(\omega_j)e^{-i(\omega_j+\Delta_{j,k}) t} \otimes \hat{B}(t),
\end{split}
\end{align}
where we denote by $e^{i\hat{K}t/\hbar}\hat{C}(\omega_j)e^{-i\hat{K}t/\hbar} = \sum_{k} \hat{C}^{(k)}(\omega_{j})e^{-i\Delta_{j,k}t}$. $\Delta_{j,k}$ is the dressing of the collapse frequencies by the effective Hamiltonian, while $\hat{B}$ has units of frequency. This is the same as the system-bath Hamiltonian expressed in the interaction picture with respect to the system Hamiltonian, since the unitary time evolution operator with respect to the system Hamiltonian can be expressed as $\hat{U}(t) = e^{-\hat{S}(t)/i\hbar}e^{-i\hat{K}t/\hbar}$.\\
We follow then the standard approach~\cite{breuer_dissipative_1997}, by doing a Picard iteration and then rewriting the von Neumann equation as an integro-differential equation, and assuming that the initial density matrix obeys $\operatorname{Tr}_B\left[H^I_\textit{sB}(t), \rho(0)\right]=0$,
\begin{align}
    \dot{\rho} = \frac{-1}{\hbar^2} \int_0^t \text{d}s \left[ \hH_\textit{sB}^\textit{I}(t),\left[\hH_\textit{sB}^\textit{I}(s), \rho(s)\right]\right].
\end{align}
Under the Born approximation the density matrix is assumed to be the tensor product of the system's density matrix and the bath vacuum $\rho(
t)=\rho_S (t) \otimes \rho_B^0$ for all $t$. Under the Markov approximation the derivative of the density matrix depends only on its value at the same time representing that the bath correlation function decays exponentially on a timescale that is small compared to the timescale on which $\rho_S(t)$ varies. By the same Markov approximation we extend the upper limit of integration to infinity. After tracing out the bath, we obtain,
\begin{align}
    \dot{\rho}_S &= \int_0^\infty \text{d}s \Tr_B\left(\left[ \hH_\textit{sB}^\textit{I}(t),\left[\hH_\textit{sB}^\textit{I}(t-s), \rho_S(t)\otimes \rho_B^0\right]\right]\right).
\end{align}
Focusing only on the first term in this nested commutator we derive the corresponding term in the Lindblad master equation, while the other terms can be treated similarly.
\begin{align}
    \begin{split}
        \dot{\rho}_S \supset &-\int_0^\infty \text{d}s \hat{C}^{(k)}(\omega_i) \hat{C}^{(k^\prime)}(\omega_j) \rho_S(t)e^{i(\omega_j+\Delta_{j,k^\prime}) s}  \\
        &e^{-i(\omega_i+\omega_j+\Delta_{i,k}+\Delta_{j,k^\prime})t}\Tr_B\left(\hat{B}(t)\hat{B}(t-s) \rho_B^0\right),
    \end{split}
\end{align}
where we used the notation $\supset$ to highlight that we are only considering one term in the nested commutator. We can then use the Onsager symmetry that states that $\Tr_B\left(\hat{B}(t)\hat{B}(t-s) \rho_B^0\right)$ only depends on $s$. We define the unilateral power spectral density of the noise
\begin{align}
    \begin{split} \label{eq:sbb}
        s(\omega)= \int_{0}^{\infty} d \tau e^{i \omega \tau} \operatorname{Tr}[ \hat{B}(\tau) \hat{B}(0) \hat{\rho}_B^0].
    \end{split}
\end{align} 
With these assumptions we obtain the following master equation,
\begin{widetext}
\begin{align}
\begin{split}
    \dot{\rho}_S = \sum_{i,j,k,k^\prime} -e^{-i(\omega_i+\omega_j+\Delta_{i,k}+\Delta_{j,k^\prime})t} &\Big\{\left[\hat{C}^{(k)}(\omega_i) \hat{C}^{(k^\prime)}(\omega_j) \rho_S(t) - \hat{C}^{(k^\prime)}(\omega_j) \rho_S(t)\hat{C}^{(k)}(\omega_i) \right]  s(\omega_j+\Delta_{j,k^\prime}) \\
    &+ \left[\hat{C}^{(k^\prime)}(\omega_j) \hat{C}^{(k)}(\omega_i) \rho_S(t) - \hat{C}^{(k)}(\omega_i) \rho_S(t)\hat{C}^{(k^\prime)}(\omega_j) \right]  s^*(\omega_j+\Delta_{j,k^\prime})  \Big\}.
\end{split}
\end{align}
\end{widetext}

Now, we make a partial secular approximation where we keep only contributions that satisfy 
\begin{align}
\omega_i+\Delta_{i,k}+\omega_j+\Delta_{j,k^\prime} < \kappa_b. 
\end{align}
The adiabaticity condition dictates $g_2 <\kappa_b$ by\cref{eq:adiabatic_condition}, hence the energy dressing coming from the effective Hamiltonian  $\Delta_{i,k}+\Delta_{j,k^\prime}$ are small compared to $\kappa_b$. Hence, after the partial secular approximation only terms with $\omega_i = -\omega_j$ are kept. We further assume that the power spectral density of the noise $s(\omega_j)$ is locally flat around any relevant frequency, i.e. $s(\omega_j) \simeq s(\omega_j + \Delta_{j,k^\prime})$ for all $j,k^\prime$ appearing in the sums above. We recast the sum over $k,k^\prime$ in terms of $e^{i\hat{K}t/\hbar} C(\omega_i) e^{-i\hat{K}t/\hbar}$ and we obtain the following Lindblad master equation,
\begin{align}
    \begin{split}
        \dot{\rho}_S &= \frac{1}{i\hbar} \left[\sum_{i} p(\omega_i)e^{i\hat{K}t/\hbar}\hat{C}^\dagger(\omega_i)\hat{C}(\omega_i)e^{-i\hat{K}t/\hbar} ,\rho_S(t)\right]\\
        & + \sum_i \kappa(\omega_i) \mathcal{D}_{e^{i\hat{K}t/\hbar}\hat{C}(\omega_i)e^{-i\hat{K}t/\hbar}}[\rho_S(t)],
    \end{split}
\end{align}
where $\kappa(\omega) = s(\omega)+s^*(\omega)$ the bilateral power spectral density of the noise and $p(\omega) =\Im[s(\omega)]$. In the present study we ignore the effect of the Lamb shift by putting the imaginary part of the spectral function to 0, $p(\omega) \to 0$. To recover \cref{eq:L_eff}, we undo the interaction picture with respect to $\hat{K}$.

\section{Kubo formula for the Floquet-Markov Liouvillian}
\label{app:abs_spect}
In this Appendix, we derive the formula for the impedance associated with a system evolving under a Floquet-Markov Liouvillian, as used in the main text. This Appendix is organized as follows. In \cref{app:impedance_summary}, we summarize the main result. The interested reader can find the derivation in the remaining subsections. In \cref{Sec:KuboTI}, we overview the Kubo formula for Lindblad evolution perturbed by a probe. \Cref{sec:Fl_Lind_Sambe} expresses the Floquet Lindbladian in the Sambe space and diagonalizes it. \Cref{sec:Kubo_Sambe} covers our derivation for the Kubo formula for a Lindbladian corresponding to a Floquet-Markov master equation.

\subsection{Impedance of a circuit in the Floquet-Markov formalism}
\label{app:impedance_summary}
In this section, we summarize the main result for the impedance of a circuit in the Floquet formalism and highlight its relation to the reflection coefficient.
The reflection coefficient is defined by
\begin{align}
    \Gamma = \frac{V_-}{V_+} = \frac{Z-Z_0}{Z+Z_0},
\end{align}
with $V_-$ the voltage of the reflected wave and $V_+$ the one input wave, $Z_0$ is the impedance of the drive line, and $Z$ of the impedance of the part of the circuit behind the drive line. The latter impedance can be further separated into the impedance of the coupling capacity between the drive line and circuit, $1/jC_g\omega$ with $C_g$ the gate capacitance, and the impedance of the circuit, $Z_S(\omega)$. We follow~\cite{pietikainen_observation_2017} but provide explicit derivations, which, to our knowledge, are not available in the literature to date. This allows us to detail the derivation of the impedance plotted in \cref{sec:strong-drive}.

One can write the impedance of a circuit $Z_S$ in terms of response functions~\cite{fluctuations_quantiques_1997}
\begin{align} \label{eq:Zs}
    Z_S[\omega]  = \chi_{\Phi V}[\omega] = \left(\frac{2e n_{zpf}}{C_g}\right)^2 \frac{1}{i\omega} \chi_{\hat{y}_b,\hat{y}_b}[\omega],
\end{align}
with $\chi_{BA}$ the response function on observable $\hat{B}$ after a perturbation proportional to operator $\hat{A}$, to be defined below, $\hat{\Phi}$ is the flux-node at node $4$, $\hat{V}$ the voltage at node $4$ [see \cref{fig:GalvaCat_node}], $e$ the charge of the electron, $C_g$ the gate capacitance and $n_{zpf}$ the zero-point fluctuation of the Cooper-pair number. The above equality was obtained by using the relation between flux and voltage, $\hat{\Phi} i\omega= \hat{V}$ and the relation between the voltage operator and the dimensionless Cooper-pair number operator $\hat{V} = \frac{2en_{\textit{zpf}}}{C_g} \hat{y}_b$. We considered only the flux and voltage variable at the node $4$ since the transmission line is connected to this port [see \cref{fig:GalvaCat}]. Moreover, we ignored the charge hybridization coefficient when expressing $\hat{V}$ as a function of $\hat{y}_b$, which is consistent with \cref{sec:strong-drive}.

The impedance of the system, provided it is prepared in an initial state corresponding to its Floquet mode $i$, to be defined below, is
\begin{align}\label{eq:Impedance}
    \begin{split}
        Z_S(\omega) \propto \frac{1}{\omega} \sum_{j,k} \left|y_{ijk}\right|^2 \Bigg[& \frac{1}{\Gamma^{(i,j)} + i(\omega -\Delta_{i,j,k})}  \\ &- \frac{1}{\Gamma^{(i,j)} + i(\omega+ \Delta_{i,j,k})} \Bigg],\\
    \end{split}
\end{align}
with $\Gamma^{(i,j)}$ defined below in \cref{eq:impedance_final}.

\begin{figure*}[t!]
    \centering
    \includegraphics[width=\textwidth]{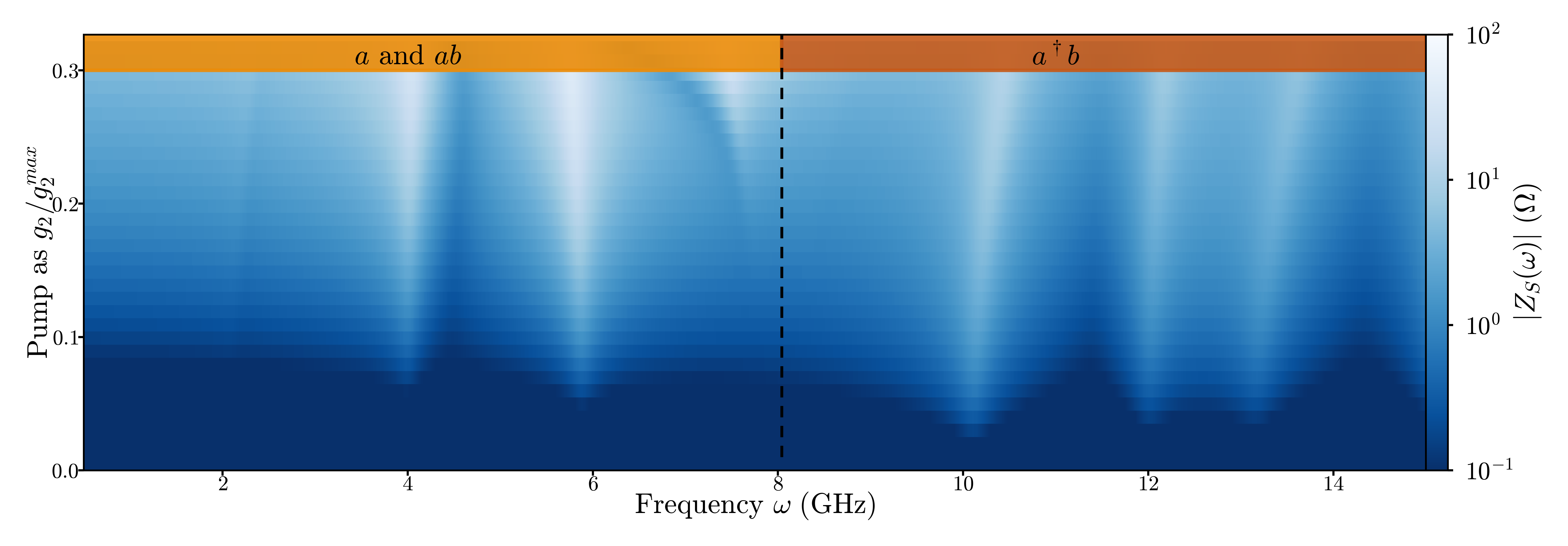}
    \caption{Impedance of the system in the Floquet state $\ket{00}$. Using \cref{eq:Impedance} with the initial state in the 0-th Floquet state, the probe frequency in swept, and the absolute value of the impedance highlights pump-activated transitions. To increase the contrast and the visibility of the plot, the transition at $\omega_d= \tilde{\omega}_b$ (i.e. $00 \to 01$ and $00\to 20$) have been subtracted. We can further attribute the spectroscopic traces to the decay processes analyzed in \cref{fig:Collapse_monom}. This plot validates the ordering of the decay processes shown in \cref{fig:Collapse_monom}. The same parameters as in \cref{sec:strong-drive} were used and $\kappa_b/2\pi = 100~\text{MHz}$.}
    \label{fig:spectroscopy_total}
\end{figure*}

In \cref{sec:strong-drive}, we further introduce the partial impedance as the contribution to the impedance stemming from a given transition $i\to j$. This amounts to separating the $j$ contributions in the above sum. This quantity is then closely related to the frequencies of the collapse operators containing operators relating state $i$ to $j$ (see \cref{sec:strong-drive}). In \cref{fig:spectroscopy_total}, we plot the impedance for the initial state in the Floquet mode labeled with the vacuum (see \cref{sec:strong-drive}).

\subsection{Kubo formula}
\label{Sec:KuboTI}
In Kubo's original formalism~\cite{kubo_statistical-mechanical_1957,giuliani_vignale_2005}, a system is perturbed by a time-dependent external perturbation. With indices $0$ for the system and $p$ for the probe, this is
\begin{align}
    \mathcal{L} = \mathcal{L}_0 + \mathcal{L}_p, 
    \label{Eq:L}
\end{align}
such that the Lindblad equation reads
\begin{align}
    \dot \rho = \mathcal{L} \rho,
\end{align}
with an unperturbed Lindblad evolution $\mathcal{L}_0$, and a time-dependent probe 
\begin{align}
    \mathcal{L}_p \cdot =  \frac{1}{i \hbar}[- A K(t), \cdot]
\end{align}
where $A$ is an operator, and $K(t)$ is a function of time. 

Assuming that the system starts in some initial state $\rho_0(-\infty)$, we wish to determine the time-dependent change in the density matrix, under the action of the perturbation
\begin{align}
    \rho_0(t) = \rho_0(-\infty) + \Delta \rho(t).
\end{align}
This is, up to linear order in the probe Lindbladian,
\begin{align}
    \Delta \rho(t)=\int_{-\infty}^t d t^{\prime} \exp \left\{\left(t-t^{\prime}\right) \mathcal{L}_0 \right\} \mathcal{L}_{p}\left(t^{\prime}\right) \rho(-\infty).
\end{align}
The response of the system is then observed by measuring another system observable $B$. We define the change in this observable as 
\begin{align} \label{Eq:Response}
    \begin{split}
    \Delta \langle B \rangle(t) &= \mathrm{Tr} \left\{ B \Delta \rho(t)  \right\} = \int_{-\infty}^t dt^\prime K(t^\prime) \chi_{BA}(t-t^\prime),
    \end{split}
\end{align} 
in terms of the response function
\begin{align}
    \begin{split}
    \chi_{BA}(t) &= \frac{1}{i\hbar}\mathrm{Tr} \left\{  [\rho_0(-\infty), A] B(t)  \right\},
    \end{split}
\end{align}
with Heisenberg-picture observable $B(t) = e^{\mathcal{L}_0^+ t } B$, with $\mathcal{L}_0^+$ the adjoint  Lindbladian~\cite{lindblad_generators_1976}.  The Fourier transform of the response function is the susceptibility that enters the definition of the impedance \cref{eq:Zs},
\begin{align}
    \begin{split} \label{Eq:chiBA}
    \chi_{BA}[\omega] &\equiv \lim_{\epsilon \to 0^+} \int_0^\infty \chi_{BA}(t) e^{-i \omega t - \epsilon t} dt.
    \end{split}
\end{align}
Note how the response is dependent on the initial state of the system $\rho_0(-\infty)$.
A number of references deal with obtaining the susceptibility when $\mathcal{L}_0$ does not correspond to unitary evolution~\cite{avron_quantum_2011,avron_adiabatic_2012,ban_linear_2015,ban_linear_2015-2,ban_linear_2017,uchiyama_master_2009,uchiyama_role_2010,uchiyama_exploring_2012,camposvenuti_dynamical_2016,chetrite_quantum_2011,saeki_comparison_2010}. Here we follow~\cite{shen_non-markovian_2017} to obtain a Kubo formula for a Floquet-Markov master equation~\cite{bluemel_dynamical_1991}. The main difficulty is that the unperturbed Lindbladian is time-dependent. In the next section, we enlarge the Hilbert space to the so-called Sambe space~\cite{sambe_steady_1973} in order to reduce the problem of evaluating the susceptibility to a time-independent problem, which will allow us to use the Kubo formalism as presented in this subsection.

\subsection{Floquet Lindbladian in Sambe space}\label{sec:Fl_Lind_Sambe}
The Floquet-Markov master equation is formulated in terms of the Lindbladian~\cite{bluemel_dynamical_1991, grifoni_driven_1998},
\begin{align}\label{eq:Blumel_Lind}
    \mathcal{L}_0 \rho = \frac{1}{i\hbar} \left[\hH_s(t), \rho\right] + \sum_{\alpha,\beta} \Gamma_{\alpha\to\beta}^{(F)} \mathcal{D}_{\ket{\phi_\beta(t)}\bra{\phi_\alpha(t)}} (\rho), 
\end{align}
with the same notations as in \cref{sec:strong-drive}.
In this subsection, we embed the Floquet Lindbladian \cref{eq:Blumel_Lind} in an enlarged Hilbert space, the Sambe space~\cite{sambe_steady_1973}, so to make it time-independent. According to the Floquet theorem, the solution to the time evolution of the density matrix is of the form
\begin{align}\label{eq:Floquet_thm}
    \hat{\rho}(t) = f(t) \hat{\sigma}(t),   
\end{align}
where $\hat{\sigma}(t+T) = \hat{\sigma}(t)$ is $T=\frac{2\pi}{\omega_d}$-periodic. We introduce the Sambe space obtained by taking the product of the system Hilbert space $\mathcal{H}$ with $\mathcal{T}$, the space of $T$-periodic functions~\cite{sambe_steady_1973, shirley_solution_1965}. We choose the orthonormal basis of $\mathcal{T}$ to be the set of functions $\left(t \to e^{ik\omega_dt} \right)_{k\in\mathbb{Z}}$. We will use the Dirac notation $\ket{k}$ to denote these quantities. We introduce an orthonormal basis of this enlarged Hilbert space using the Floquet modes of the problem $\ket{\phi_i(t)}$. To this end, we denote $\ket{j,k} = t\to e^{ik\omega_d t}\ket{\phi_j(t)}$,
\begin{align}
    \ket{j,k} = \sum_l \int_0^{T} \ket{\phi_j(t)} e^{i((k-l)\omega_d t } \frac{\text{d}t}{T} \otimes \ket{l},
\end{align}
here we have projected $\ket{j,k}$ on the basis $\ket{k}$ of $\mathcal{T}$ using the inner-product of two periodic functions $\braket{f}{g}=\int_0^T f^* g \frac{\text{d}t}{T}$.
The orthonormality of the $\ket{j,k}$'s can be proven using $\sum_l e^{il\omega_d (t-t^\prime)} = T \delta(t-t^\prime)$.

Similarly, we view a time-periodic operator $\hat{O}(t)$ as a linear map on $\mathcal{H}\otimes \mathcal{T}$ which to an element $t\to \ket{\phi(t)}$ associates $t\to \hat{O}(t) \ket{\phi(t)}$. To avoid confusions we call this map $\mathcal{O}$, we can decompose this map on the $k$-basis,
\begin{align}\label{eq:decomp_O}
    \mathcal{O} = \sum_{n,k} \int_0^T \hat{O}(t) e^{i(k-n)\omega_dt}\frac{\text{d}t}{T}\otimes \ket{n}\bra{k}.
\end{align}
We derive the action of the $\partial_t$ operator in this enlarged Hilbert space. Using the Floquet theorem, the set of functions on which $\partial_t$ acts is given by \cref{eq:Floquet_thm}.
We have,
\begin{align}
    \partial_t f \hat{\sigma} = \dot{f} \hat{\sigma} + f \dot{\hat{\sigma}}. 
\end{align}
We decompose the linear map on $\mathcal{H}\otimes\mathcal{T}$, $\partial_t{\varsigma}$, associated to $\dot{\hat{\sigma}}(t)$ in the $k$-basis,
\begin{align}
    \partial_t{\varsigma} = \sum_{n,k} \int_0^T i (n-k)\omega_d \hat{\sigma}(t) e^{i(k-n)\omega_dt}\frac{\text{d}t}{T}\otimes \ket{n}\bra{k}.
\end{align}
We see that $\partial_t\hat{\mu} \otimes \ket{n}\bra{k} = i(n-k)\omega_d~ \hat{\mu}\otimes\ket{n}\bra{k}$, with $\hat{\mu}$ a time-independent operator on $\mathcal{H}$.
Finally, the action of $\partial_t$ on a test element $f ~\hat{\mu}\otimes \ket{n}\bra{k}$, where $f$ is not necessarily a periodic function, 
\begin{align}\label{eq:Sambe_derivative}
\begin{split}
    \partial_t (f~ \hat{\mu} \otimes \ket{n}\bra{k}) &= \dot{f}~\hat{\mu} \otimes \ket{n}\bra{k} \\
    &+ i\omega_d(n-k)f~\hat{\mu} \otimes \ket{n}\bra{k}\\
    \partial_t(\cdot)  &= \partial_t^{\mathcal{H}\otimes\mathcal{T}_{\omega_d}}(\cdot) +i \left[\sum_n \mathbb{I} \otimes n\omega_d \ket{n}\bra{n}, \cdot \right],
\end{split}
\end{align}
where we have introduced the notation $\partial_t^{\mathcal{H}\otimes\mathcal{T}_{\omega_d}}$ to emphasize that the time-derivative acts only on the non-periodic part. 

Before expressing the full Lindbladian, we calculate the action of the $\hH_s(t) -i\hbar\partial_t $ on our basis $\ket{j,k}$.
By definition of the Floquet modes we have $\left(\hH_s(t) -i\hbar\partial_t \right)e^{ik\omega_dt} \ket{\phi_j(t)} = \hbar(\epsilon_j + k\omega_d) e^{ik\omega_dt} \ket{\phi_j(t)}$. Therefore
\begin{align}
\begin{split}
    &\left[\hH_s(t) -i\hbar \partial_t \right]e^{ik \omega_d t} \ket{\phi_j(t)} \\
    &=\left[\hH_s(t)+\sum_n \mathbb{I} \otimes n\hbar\omega_d \ket{n}\bra{n}\right] \ket{j,k} \\
    &= \hbar(\epsilon_j + k\omega_d) \ket{j,k},
\end{split}
\end{align}
and then
\begin{align}
\begin{split}
    \hH_s(t)-\sum_n \mathbb{I} \otimes n\hbar \omega_d \ket{n}\bra{n} 
    &= \sum_{j,k}\hbar(\epsilon_j+k\omega_d)\ket{j,k}\bra{j,k}.
\end{split}
\end{align}

Using the equation \cref{eq:decomp_O} to express $\ket{\phi_\beta(t)}\bra{\phi_\alpha(t)}$ in the $k$-basis and calculating the action on $\ket{i,k}$ one finds,
\begin{align}
     \left(\ket{\phi_\beta(t)} \bra{\phi_\alpha(t)}\right) \ket{i,k} =\delta_{i,\alpha}\ket{\beta,k}.
\end{align}
Note that this differs from $\ket{\beta,0} \braket{\alpha,0}{i,k}= \delta_{i,\alpha} \delta_{k,0} \ket{\beta,0}$. This can be understood by the fact that $\ket{\phi_\beta(t)}\bra{\phi_\alpha(t)} \neq \ket{\beta,0} \bra{\alpha,0}$. So we find that $\ket{\phi_\beta(t)} \bra{\phi_\alpha(t)} = \sum_k \ket{\beta,k}\bra{\alpha,k}$.

Finally, we can write the Floquet Lindbladian as a time-independent superoperator in the Hilbert space $\mathcal{H}\otimes \mathcal{T}_{\omega_d}$. We denote by $\varrho$ a density matrix of the form given by the Floquet theorem \cref{eq:Floquet_thm}. In the $\ket{i,k}$ basis we have,
\begin{align}
    \mathcal{L}^{\mathcal{H}\otimes \mathcal{T}_{\omega_d}}(\varrho) &= \frac{1}{i}\left[\sum_{j,k}(\epsilon_j+k\omega_d)\ket{j,k}\bra{j,k}, \varrho\right]\\
    &+ \sum_{\alpha,\beta} \Gamma_{\alpha\to \beta}^{(F)} \mathcal{D}_{\sum_k \ket{\beta,k}\bra{\alpha,k}}(\varrho),
\end{align}
where the part of the action of the time-derivative that is linear has been incorporated in the definition of the Lindbladian in Sambe space. In Sambe space, we have a time-independent Lindbladian evolution
\begin{align}
    \partial_t^{\mathcal{H}\otimes\mathcal{T}_{\omega_d}}\varrho &= \mathcal{L}^{\mathcal{H}\otimes\mathcal{T}_{\omega_d}}(\varrho).
\end{align}
To calculate the evolution operator $e^{t\mathcal{L}}$ we further diagonalize the Lindbladian in Sambe space. The eigenelements of the Floquet Lindbladian are for $i\neq j$,
\begin{align}
    \mathcal{L}^{\mathcal{H}\otimes \mathcal{T}_{\omega_d}}(\ket{i,k}\bra{j,l}) &= \left[-i\Delta_{ij,k-l} -\Gamma^{(i,j)}\right]\ket{i,k}\bra{j,l} ,
\end{align}
where $\Gamma^{(i,j)} = \frac{1}{2}\sum_\beta ( \Gamma_{i\to\beta}^{(F)}+ \Gamma_{j\to\beta}^{(F)})$. 

The remaining eigenelements are obtained, by diagonalizing the matrix $M_{i,j} =\Gamma_{j\to i}^{(F)} - \delta_{i,j}\Gamma_{i\to i}^{(F)}$. Denoting the eigenvector $p_i^{(n)}$ associated to the $n$-th eigenvalue $\gamma^{(n)}$, such that $\sum_i M_{i,j} p_i^{(n)} = \gamma^{(n)} p_j$. We find the remaining eigenelements of the Floquet Lindbladian,
\begin{align*}
    \mathcal{L}^{\mathcal{H}\otimes \mathcal{T}_{\omega_d}}&\left(\sum_i p_i^{(n)}\ket{i,k}\bra{i,l} \right)= \\
    &\left[-i (k-l)\hbar\omega_d +\gamma^{(n)}  \right]\sum_i p_i^{(n)}\ket{i,k}\bra{i,l}. 
\end{align*}
We extend the previous notation to include this case with $\Delta_{iik} = \hbar k\omega_d$.

\subsection{Kubo formula in Sambe space}\label{sec:Kubo_Sambe}
The Kubo formula readily extends to Sambe space using the derivation in \cref{Sec:KuboTI}. The main problem is to identify the `initial' condition. We choose $\rho_0(-\infty) = \ket{i,0}\bra{i,0}$, meaning that the system starts in a Floquet eigenmode. Note that this is a time-periodic density-matrix and does not correspond to an initial condition. However, it seems to be a natural choice in a Floquet problem to consider perturbations on an initial state in a Floquet mode. The impedance can then be recast as,
\begin{align}
\begin{split}
    \chi_{\hat{y}_b,\hat{y}_b}(t) &= \frac{1}{i\hbar} \mathrm{Tr} \left[e^{t\mathcal{L}^{\mathcal{H}\otimes \mathcal{T}_{\omega_d}}} \left(\left[\ket{i,0}\bra{i,0}, \hat{y}_b\right] \right) \hat{y}_b\right],\\
    & = \sum_{j,l}\frac{1}{i\hbar} \bra{i,0}\hat{y}_b\ket{j,l} \\
    &\mathrm{Tr} \left[ e^{t\mathcal{L}^{\mathcal{H}\otimes \mathcal{T}_{\omega_d}}} \left(\ket{i,0}\bra{j,l} - \ket{j,l}\bra{i,0} \right) \hat{y}_b \right],
\end{split}
\end{align}
We finally make the approximation\\ 
\begin{align}
\mathcal{L}^{\mathcal{H}\otimes \mathcal{T}_{\omega_d}} \left( \ket{i,0}\bra{i,l}\right) \simeq \left[-i\Delta_{ii,-l} -\Gamma^{(i,i)}\right] \ket{i,0}\bra{i,l},
\end{align} 
where $\Gamma^{(i,i)} = \sum_{j\neq i} \Gamma_{j\to i}^{(F)}$. This approximation amounts to discarding the off-diagonal contributions of $M_{i,j}$.

By further noticing that $\bra{i,0}\hat{y}_b\ket{j,l} = y_{ij,-l}$ (see \cref{sec:strong-drive}) we get,
\begin{align}
    \chi_{\hat{y}_b,\hat{y}_b}(t)& = \sum_{j, l}\frac{\left|y_{ij,-l}\right|^2}{i\hbar}    \left(e^{-i\Delta_{ij,-l}t -\Gamma^{(i,j)}t}- e^{-i\Delta_{jil}t -\Gamma^{(i,j)}t} \right),
\end{align}
which finally results in the expression for the impedance
\begin{align}\label{eq:impedance_final}
\begin{split}
    Z_S(\omega) \propto \frac{1}{\omega} \sum_{j,k} \left|y_{ijk}\right|^2 \Bigg[& \frac{1}{\Gamma^{(i,j)} + i(\omega -\Delta_{ijk})}  \\ &- \frac{1}{\Gamma^{(i,j)} + i(\omega+ \Delta_{ijk})} \Bigg],\\
    \Gamma^{(i,j)} =\sum_n \frac{\Gamma_{n\to i}^{(F)} + \Gamma_{n\to j}^{(F)}}{2}  &- \delta_{i,j} \Gamma_{i\to i}^{(F)}.
\end{split}
\end{align}

\bibliography{ref}

\end{document}